\documentclass[acmlarge, screen, nonacm]{acmart}

\usepackage{enumitem}
\usepackage[]{mdframed}
\usepackage{caption}
\usepackage{subcaption}
\usepackage{listings}
\usepackage{booktabs}
\usepackage{multirow}

\setcopyright{none}
\renewcommand\footnotetextcopyrightpermission[1]{}
\newcommand{\squishlist}{\begin{itemize}[itemsep=1pt,parsep=2pt,topsep=3pt,partopsep=0pt,leftmargin=0em, itemindent=1em,labelwidth=1em,labelsep=0.5em]}
\newcommand{\squishend}{\end{itemize}}

\AtBeginDocument{%
  \providecommand\BibTeX{{%
    \normalfont B\kern-0.5em{\scshape i\kern-0.25em b}\kern-0.8em\TeX}}}

\begin{document}

\title{SnappyMeal: Design and Longitudinal Evaluation of a Multimodal AI Food Logging Application}

\author{Liam Bakar}
\email{lbakar@uw.edu}
\author{Zachary Englhardt}
\email{zacharye@cs.washington.edu}
\author{Vidya Srinivas}
\email{vysri@cs.washington.edu}
\author{Girish Narayanswamy}
\email{girishvn@uw.edu}
\author{Dilini Nissanka}
\author{Shwetak Patel}
\email{shwetak@uw.edu}
\author{Vikram Iyer}
\email{vsiyer@uw.edu}
\affiliation{
  \institution{University of Washington}
  \city{Seattle}
  \state{Washington}
  \country{USA}
}

\renewcommand{\shortauthors}{Bakar et al.}

\begin{abstract}
Food logging, both self-directed and prescribed, plays a critical role in uncovering correlations between diet, medical, fitness, and health outcomes. 
Through conversations with nutritional experts and individuals who practice dietary tracking, we find current logging methods, such as handwritten and app-based journaling, are inflexible and result in low adherence and potentially inaccurate nutritional summaries. These findings, corroborated by prior literature, emphasize the urgent need for improved food logging methods. 
In response, we propose SnappyMeal, an AI-powered dietary tracking system that leverages multimodal inputs to enable users to more flexibly log their food intake.
SnappyMeal introduces goal-dependent follow-up questions to intelligently seek missing context from the user and information retrieval from user grocery receipts and nutritional databases to improve accuracy. 
We evaluate SnappyMeal through publicly available nutrition benchmarks and a multi-user, 3-week, in-the-wild deployment capturing over 500 logged food instances. 
Users strongly praised the multiple available input methods and reported a strong perceived accuracy. 
These insights suggest that multimodal AI systems can be leveraged to significantly improve dietary tracking flexibility and context-awareness, laying the groundwork for a new class of intelligent self-tracking applications.

\end{abstract}

\maketitle

\begin{figure*}[ht!]
    \centering
    \includegraphics[width=.75\linewidth]{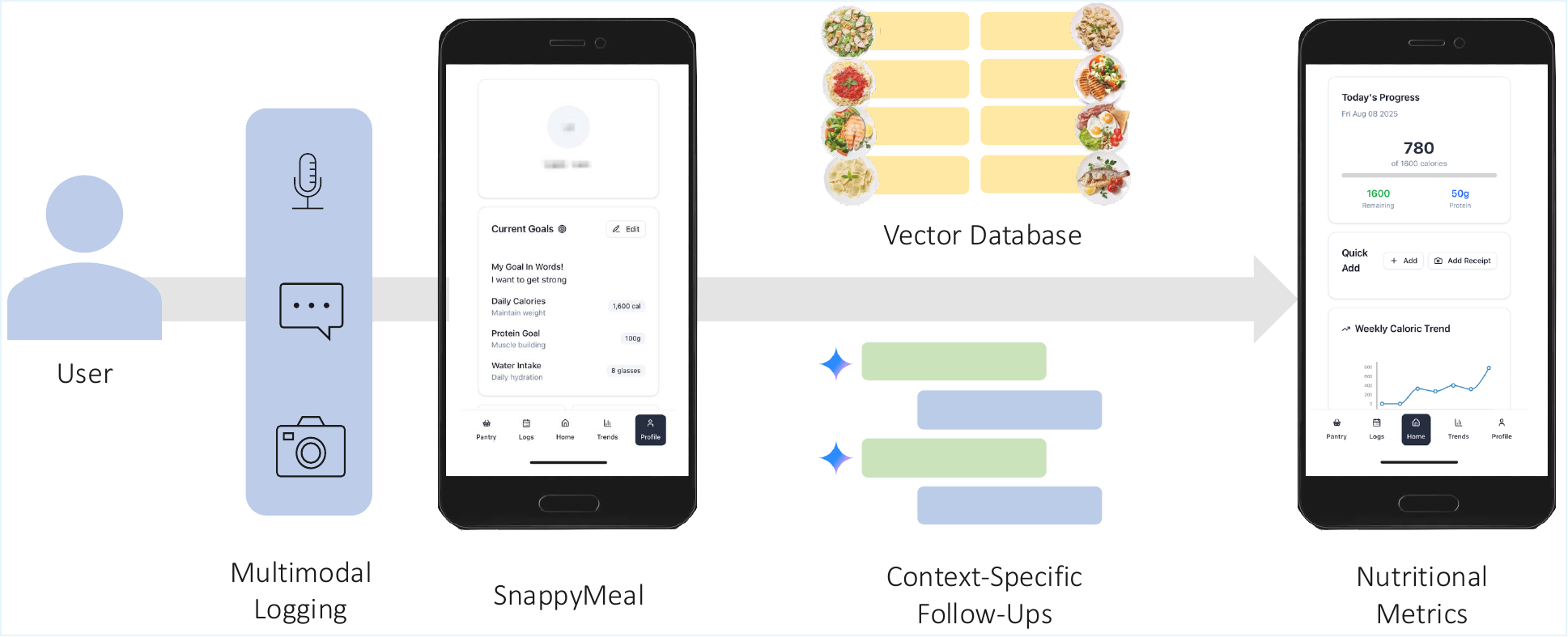}
    \caption{Traditional food logging using handwritten diaries or even mobile applications rely heavily on manual data entry or suffer from poor accuracy estimation techniques. We instead develop a smartphone-based multimodal AI system that combines diverse multimodal context from food and receipt images to natural language text and audio with interactive follow up questions to improve tracking flexibility and contextual awareness.}
    \label{fig:placeholder}
    \Description[Demonstrating antiquated food logging against novel food logging.]{Demonstrating antiquated food logging, using scales, calculators, and notebooks, against novel food logging, using mobile devices, multiple modalities, and artificial intelligence.}
\end{figure*}

\section{Introduction}

It is well studied that proper nutrition and eating habits are correlated with healthier living and reduced risk for a number of conditions~\cite{kandel2019an}. However, what constitutes healthy diet and dietary patterns varies greatly from individual to individual, and is influenced by factors such cultural and demographic background, geographic location, pre-existing conditions, and  food intolerance/sensitivities. The definition of healthy diet is further convoluted by factors such as self-image, societal pressures, socioeconomic status, and personal goals. As such there is not a “one-size-fits-all” solution to individual eating habits. To this end, food logging, both self-directed and prescribed, plays a critical role in uncovering correlations between diet, medical, fitness, and health outcomes for individuals~\cite{dahake2025a,verma2018challenges}.

% To this end, food logging, both self-directed and prescribed, plays a critical role in uncovering correlations between diet, medical, fitness, and health outcomes.
% In recent years, nutrition has taken center stage in public health conversations, with approximately 40\% of adults actively trying to lose weight~\cite{bleich2014patient}.
% At the same time, 2–4\% of young adults struggle with clinically diagnosed eating disorders~\cite{swanson2011prevalence}, underscoring the fine line between health-conscious behavior and disordered eating.
% Eating behavior is further complicated by the fact that food is not only essential for survival but also deeply embedded in cultural identity and social relationships, meaning individuals often lack complete control over what, when, and how they eat. Tracking food plays a critical role in diagnosing and managing a wide variety of diet-related health conditions and fitness goals~\cite{dahake2025a,verma2018challenges}.

Nutritional tracking tools have emerged as a way to bring more awareness and structure to eating habits. These tools enable the 69\% of U.S. adults keeping track of at least one health indicator (weight, diet, exercise regimen, or symptom)~\cite{fox2013tracking} to better monitor intake, set goals, and reflect on patterns using smartphone and wearable based logging. However, unlike fitness trackers (e.g., smartwatches) that can automatically measure health metrics (e.g., step counts, heart rate), 
modern nutritional tracking tools, both handwritten and app-based, largely depend on manual inputs.

To better understand the current landscape of nutrition tracking apps, we conduct a formative study with nine participants spanning both professional dietitians and food journalers (those who log diet). Our results show that usability and convenience, psychological impact, long-term sustainability, and the risk of tracking fatigue (a common phenomenon in which users disengage due to the effort and mental load of constant self-monitoring~\cite{kyoung2014understanding}) all contribute to low adherence and tracking accuracy~\cite{todd1983food, griffiths2018assessment}. Notably, users seek intuitive and flexible systems that align better with their personalized goals and lifestyle.

% Based on these findings we develop \emph{SnappyMeal}, an AI-powered dietary tracking system that leverages multimodal inputs to enable users to more flexibly and accurately log their food intake. SnappyMeal lowers the barrier to tracking by allowing users to log their food using unstructured text, speech, and images inputs. We develop an AI system that automatically extracts relevant information from these diverse inputs, leverages a nutritional databases to improve accuracy, and interactively prompts users with goal-dependent follow-up questions to intelligently extract important context.

Designing a single tracking solution that will seamlessly fit into diverse users' lifestyles is challenging, as each individual has varying dietary needs and nutrition tracking objectives.
A person aiming for muscle gain has different nutritional needs (e.g., higher protein intake) than someone trying to manage cholesterol or lose weight~\cite{uauy2005defining}. 
%Personalized tracking can be fine-tuned to help an individual achieve their specific goals. 
Dietary guidelines provided by organizations such as the FDA are not specific or tailored enough to provide nutritional guidance for each individual~\cite{adams2020perspective,renner2023perspective}. While many apps exist for self tracking (MyFitnessPal, LoseIt, etc), our formative study finds they fall short when estimating portion sizes and matching food items~\cite{chen2019the, griffiths2018assessment}. Many tools do not log important context about food preparation or seasonings, and struggle on the challenging problem of estimating specific nutrients.
%Whether individuals choose to consult registered dietitians or personalized nutrition advice systems, the need for tracking remains paramount~\cite{dahake2025a,verma2018challenges}. 
%However, current tracking methods are either laborious~\cite{todd1983food} or inaccurate~\cite{griffiths2018assessment}.

Recent advances in artificial intelligence (AI) open up new opportunities to reimagine how nutrition tracking systems can be designed and experienced. In particular, multimodal foundation models can jointly process text, images, and audio, allowing users to describe meals in natural language, upload photos of food or receipts, or record brief voice notes. These capabilities enable more flexible and personalized methods of logging that may better fit diverse lifestyles, contexts, and abilities.
At first glance, such technology appears to offer a straightforward solution to the long-standing challenges of nutritional tracking. However, deploying AI-powered systems reveals that many core issues remain unsolved. Even with multimodal inputs, models often lack the contextual awareness needed to make accurate inferences: photos may be taken at poor angles, ingredients may be occluded, or voice notes may omit crucial details. The challenge, then, is how to fill in this missing context without increasing user burden or disrupting the tracking experience.

To explore this opportunity and its associated challenges, we develop 
\emph{SnappyMeal}, an AI-powered nutrition tracking application that integrates multimodal inputs, retrieval-augmented context, and adaptive interactivity. \emph{SnappyMeal} surfaces the tensions at the intersection of AI, HCI, and system design, balancing automation with user flexibility and context-seeking, and highlights that simply ``adding AI to the loop” does not automatically solve the difficult problems of nutrition tracking. Instead, meaningful progress depends on understanding what users value, how they wish to engage with AI assistance, and how systems can adapt to individual goals and contexts without imposing additional cognitive or interactional load.
We introduce three complementary strategies for context augmentation in AI-powered nutrition tracking: (1) leveraging RAG to retrieve not only structured nutritional data but also visually and semantically similar food images, enhancing contextual understanding, (2) incorporating receipts to provide additional context about what the user actually bought, reducing reliance on potentially incomplete images or descriptions, and (3) selectively generating follow-up questions that directly elicit missing information from users in a goal-directed manner. Together, these strategies aim to improve flexibility and personalization of food logging while mitigating user effort and cognitive load.

We summarize our contributions below:
\begin{itemize}
    \item We conduct a formative study identifying both the key gaps in current leading food tracking technologies and opportunities for AI systems to close these gaps.
    %VI I'm going to combine these two 
    %\item We develop an end-to-end system including mobile app and cloud system enabling users to easily provide diverse, multimodal context about their meals.
    %\item We develop the first AI system that interactively queries users to gather more context about food.
    \item We develop an end-to-end mobile and cloud system that supports diverse multimodal inputs—images, natural language text, and speech—to capture meal context, and introduce the first AI system that interactively queries users to fill in missing details.
    %\item We conduct a three-week longitudinal study, yielding over 500 food logs, to assess the challenges in incorporating flexibility and context-awareness in real-world nutrition tracking.
    \item We investigate these AI-powered features through a real-world, three-week longitudinal study yielding over 500 food logs revealing both progress and key challenges in using AI to create flexible, context-awareness nutrition tracking systems.
\end{itemize}
\section{Related Work}
\label{section:related-work}
% Our review of the prior literature begins with an overview of the benefits and challenges of tracking nutrition for health. Next, we outline technological innovations introduced to digitize existing nutrition-tracking practices. Finally, we discuss similar studies and their evaluations of such digitized systems.

\subsection{Nutrition Tracking For Health}
%VI: For the next revision I think we can shorten this to one paragraph. The way to do this is you can talk about a whole group of works that have some feature and list multiple citations

Dietary tracking can help individuals with chronic conditions better understand their health, treatments, and how these correlate with dietary intake. For example, systems like DIETOS~\cite{giuseppe2018DIETOS}, an recommendation system developed to deliver nutritional information to improve the quality of life of healthy subjects and patients with diet-related chronic diseases, promotes dietary mindfulness and through this demonstrate how personalized solutions may enhances quality of life (QoL). Similarly, dietary tracking may better inform the treatment of such chronic conditions. Several studies have shown the utility of dietary tracking in enabling diabetics to better dose insulin based on their logged carbohydrate intake~\cite{darby2016a, usman2021the}. In the same vein, another study, ~\citet{misra2025the} find that participants with type two diabetes who consistently tracked their diet improved dietary self-efficacy and intake over 6 months. Such tracking solutions have also shown utility for those in periods of convalescence, such as cancer survivors. Specifically, \citet{wang2022mHealth} find that dietary tracking drives significant positive behavioral change and improved QoL for 11 out of 18 individuals. Finally, \citet{senthilkumar2024stay} attest that dietary tracking is more than a tool for observation. Their study demonstrates that personalized dietary counseling and reminders result in higher dietary adherence, increased QoL, and improved metabolic health.

\subsection{Technology in Nutrition Tracking}

Emergent technologies are rapidly transforming the landscape of nutrition tracking, allowing users to move beyond the traditional self-report methods to incorporate automated, continuous, and objective data streams. These tracking approaches can be largely categorized as those which derive additional insights from either body-worn sensors (e.g., wearables) or from ubiquitous mobile devices.

\subsubsection{Wearables}

Several studies have explored the use of wearable devices for automated dietary assessment. \citet{amft2005analysis} utilized wearable microphones to classify food type and quantity based on chewing sounds. Similarly, other researchers have developed specialized wearable sensors for a similar purpose, such as the neckband created by \citet{cheng2013activity} to detect swallowing motions. More recently, \citet{rescic2019mobile} demonstrated the potential of gesture recognition from wrist-worn devices to quantify food intake. Furthermore, \citet{mirtchouk2016automated} developed a multi-sensor approach, combining in-ear audio sensors with head and wrist motion detectors to classify consumption. More recently, the commoditization of continuous glucose monitors (CGM) has enabled the tracking of the metabolic responses directly from the bloodstreams~\cite{ameen2025non, sunstrum2025wearable}. 
% Beyond motion detection, advances in continuous glucose monitors (CGM) are allowing nutrients to be detected directly from bloodstreams~\cite{ameen2025non, sunstrum2025wearable}. 
While promising, these methods largely rely on specialized hardware. In contrast, our work focuses on leveraging ubiquitous mobile phones and their sensing capabilities to provide a more convenient and unobtrusive method for nutritional tracking.

% While these methods show great promise for continuous, background sensing of dietary habits, they often rely on specialized hardware that may not be widely accessible or comfortable for users over long periods.
% In contrast, our work focuses on leveraging ubiquitous mobile phones and image-based analysis to provide a more convenient and unobtrusive method for nutritional tracking.

\subsubsection{Mobile Device Images}
Image-based tracking provides an unobtrusive method of logging that can be facilitated through modern mobile devices. Towards this, \citet{wang2024measure} proposed a model capable of estimating nutritional content from images of foods.
% developed a model to measure how different ingredients affect the total nutritional value of a meal, demonstrating an increase in accuracy resulting from knowing individual ingredients rather than simply holistic nutritional values.
However, images alone are often not enough to gain a comprehensive understanding of nutrition.
\citet{biel2018bites} validates that combining mobile phone image capture with contextual metadata provides a convenient and non-intrusive method for nutritional tracking.
% allowing researchers and technology developers to gain valuable insights into eating behaviors. 
The Foodprint study \cite{chung2019identifying} demonstrates that photo-based dietary diaries act as crucial "boundary negotiating artifacts," structuring data exchange to enable health experts to more quickly focus on the patient's context and specific goals.
A study conducted by \citet{shahabi2025unveiling} demonstrates the power of combining passive visual data with in-the-moment psychological and contextual data.
This combination of data can be used to predict overeating episodes and identify distinct psychological phenotypes, underscoring the importance of collecting contextual data alongside primary food logs.

\subsection{Contextual AI}

Context provides the essential information needed to resolve ambiguity and validate the significance of a measurement.
Contextual data, acquired automatically or through prompting in AI systems, transforms food log entries into actionable, accurate records. Follow-up questions can bolster contextual human-led inputs or conversations~\cite{yeomans2019it}. 
Dynamically generated follow-up questions represent a largely unsolved problem in the human-centered computing space~\cite{moore1996dynamically}. 
\citet{zhang2025harnessing}'s findings demonstrate the feasibility and effectiveness of integrating AI-generated follow-up questions
into real-time, semi-structured interviews. \citet{kuric2025unmoderated} validates the ability of models like GPT-4 to generate follow-up questions in usability testing contexts. 
In general, information elicitation tasks, supported by conversational agents, greatly benefit from follow-up question generation~\cite{hu2024designing, meng2023followupqg}.
These techniques can be extended to other domains such as clinical~\cite{li2024beyond, englhardt2024from} and nutrition~\cite{sosa2024the, chew2022the} settings.

% The advancement of image-based dietary analysis is heavily reliant on the quality and scale of available datasets. Traditional food image datasets suffer from limitations, including a small number of images, a focus solely on finished dishes, and a lack of links to contextual recipe information. To address these critical limitations, \citet{harashima2017cookpad} introduced the Cookpad Image Dataset, containing $1.64$ million images of finished food and over $3.10$ million images taken during the cooking process. This enables the development of models that can analyze a dish not just by its final appearance, but by its constituent ingredients and preparation steps—a necessary step toward highly accurate, non-intrusive dietary assessment.

\subsection{Evaluating Logging Apps}

Mobile food logging, though proven to have many benefits, remains tedious and difficult, and is a significant focus in recent literature. Some logging methods, such as selecting a meal from a large food database, can present usability challenges due to the vast amount of information crowded onto a small screen. \citet{jung2020foundations} addressed this issue through the design and evaluation of the EaT app, analyzing the timeliness of logging and identify the causes of search failures, including an analysis of 1,163 user-created entries.

In a similar vein, \citet{griffiths2018assessment} conducted a study to assess the precision of five popular free apps (including MyFitnessPal, Lose It!, and Fitbit) by comparing their nutrient intake estimates against calculations from the research-grade Nutrition Data System for Research (NDSR). From this study, it is clear that the demonstrated accuracy of automated technologies must be balanced with user compliance to ensure utility in real-world settings. The respective benefits and drawbacks of manual food journaling (high detail, high burden) and automated dietary monitoring (ADM) (low burden, lower context/detail) suggest the value of semi-automated journaling systems that combine both approaches.

\citet{lu2022understanding} address this gap by examining how people anticipate and accept these hybrid systems. Their findings establish critical design trade-offs: User satisfaction is contingent on the quality of intervention. Participants showed more positive anticipation for prompts that contained information relevant to their journaling goals, aided recall of specific foods, and did not provide too much logging burden. This validates the need for semi-automated systems to produce high-value, food-specific prompts, even if this task is ``more challenging to produce than manual reminders."
This work suggests that the true measure of a nutrition-tracking app lies not just in its technical accuracy, but in its ability to balance sensing performance with user-anticipated burdens, reinforcing the need to select tracking approaches based on individual and practitioner journaling needs.
We extend and complement this body of work studying food logging apps by specifically investigating how multimodal and conversational AI features can be incorporated into nutrition tracking apps.

\section{Formative Study}

% \textcolor{red}{VI: Looks good. I think some parts could be made more concise but I think the info is there and the message gets through. It's better to focus on other parts now, that can come in a revision.}

We performed a series of semi-structured interviews to identify strengths and shortcomings of current dietary self-tracking techniques from the perspective of both dietitians and individuals who participate in self-tracking (food journalers). Building off the insights of the literature discussed in Sec.~\ref{section:related-work}, we compiled a list of interview questions to better understand how these challenges manifest and identify potential opportunities for improvement. We focused on the nutritional tracking process itself, including frequency and consistency of tracking, methods individuals use or prescribe, how individuals interpret and use nutrition data, and the specific goals motivating dietary self-tracking. After conducting a pilot interview, we determined a semi-structured interview format would be most suitable to elicit detailed responses. 

One crucial goal of our formative study was to identify areas in which perspectives on desired improvements in tracking aids vary between dietitians and food journalers, especially relating to data accuracy and behavior change. By synthesizing these unique viewpoints, we developed a nuanced understanding of the current limitations and future opportunities for improving nutrition tracking applications and synthesized insights to guide the subsequent design of SnappyMeal. 

\subsection{Methods}

\subsubsection{Participants} We recruited 4 professional dietitians (Table ~\ref{tab:participants_dietitians}) and 5 food journalers (Table~\ref{tab:participants_loggers}) to interview on Zoom video conferencing. Dietitians were recruited through emails to School of Public Health of a major research institution, as well as other smaller departments from the research team's professional connections. Members of the general public that participate in self-tracking were recruited through word of mouth or digital and physical flyers. To aid in recruitment, participants were offered an electronic gift card, with a value of \$50 USD for dietitians and \$20 USD for members of the general public. Prior to recruiting participants, our study protocols for each population group were submitted to and approved by the IRB at the host institution for this study. 

\begin{table}[h]
    \centering
    \begin{tabular}{cccccc}
        \toprule
        \textbf{Participant ID} & \textbf{Age Range (Years)} & \textbf{Sex} & \textbf{Tracking Frequency} & \textbf{Tracking History} & \textbf{Occupation} \\
        \midrule
        L1 & 18-24 & M & Several times a week & 6 months to 1 year & Student \\
        L2 & 18-24 & M & More than once a day & 1+ years & Student \\
        L3 & 18-24 & M & More than once a day & 1-6 months & Sports Operations \\
        L4 & 18-24 & F & More than once a day & 1-6 months & Engineer  \\
        L5 & 18-24 & F & More than once a day & 6 months to 1 year & Student \\
        \bottomrule
    \end{tabular}
     \caption{Characteristics of Formative Study Participants (Food Journalers)}
    \label{tab:participants_loggers}
\end{table}

\begin{table}[h]
    \centering
    \begin{tabular}{cccc}
        \toprule
        \textbf{Participant ID} & \textbf{Years of Experience} & \textbf{Clinic Size (Number of Patients)} & \textbf{Focus}  \\
        \midrule
        D1 & 6 &  10-20 &  Weight loss and metabolic health \\
        D2 & 6 & >100 & Geriatrics \\
        D3 & 33 & 15-20 & Eating Disorders \\
        D4 & 21 & >1000 & Weight management  \\
        \bottomrule
    \end{tabular}
    \caption{Characteristics of Formative Study Participants (Dietitians)}
    \label{tab:participants_dietitians}
\end{table}

\subsubsection{Interview} We conducted a semi-structured interview using our prior knowledge of nutrition tracking as a baseline for questions, asking tailored follow-up questions and allowing participants to dive deep into their experience assisting food journalers or their experience journaling themselves. These interviews lasted approximately one hour and were recorded and transcribed asynchronously. 

\subsubsection{Analysis} Audio recordings of all interviews were transcribed using Zoom teleconferencing software. These transcripts were then subjected to an open-coding analysis. The raw audio recordings were retained as a fallback to resolve occasional transcription errors and to ensure the fidelity of the data. Subsequently, the research team performed a thematic analysis~\cite{braun2006using} to identify key insights. This analysis focused on two distinct research topics: understanding dietitians’ perspectives on how self-tracking can be improved to facilitate the promotion of healthier eating habits, and identifying self-tracking individuals' perspectives on how logging applications can be enhanced to better support self-tracking.

% https://zachary.englhardt.com/pdfs/englhardt_chi_2025.pdf

\subsection{Findings}
\label{section:formative_findings}

%VI: I added some paragraph breaks to break up the blocks of text
\subsubsection{Dietitians}
\label{section:formative_findings_dietitians}
Interviews with licensed dietitians revealed significant challenges with current dietary tracking, both manual and digital. Dietitians reported that patients, particularly older adults with limited tech skills, struggle with consistency and accuracy. As D2 stated, \textit{"Think about the least amount of work patients need to do."} Manual logs are often incomplete, missing details like portion sizes, seasonings (especially sodium), drinks, meal timing, and eating speed. Digital apps, such as MyFitnessPal and LoseIt, frequently contain inaccurate food labels, struggle to differentiate food types (e.g., steak vs. roast), and make tracking specific nutrients like fiber, sodium, calcium, and potassium difficult for many. 

Photos, while sometimes helpful, still present difficulties in accurately estimating quantities and lack context regarding preparation methods or processed versus fresh status. D2 noted, \textit{"The challenge with photos is I can’t tell how big the plate is in the image and I can't tell that there are seasonings."} Patient engagement varies, often dropping off without clear feedback or visualization of how tracking impacts their condition or relates to personal goals (e.g., energy to play with grandkids, bone health). D4 also highlighted this challenge, saying, \textit{"Patients don’t track enough and people only remember 30\% of the stuff that they track."}

Based on these challenges, dietitians expressed a clear need for improved tools. They desire simplified, low-effort tracking methods, perhaps using pre-printed templates or involving family members. Enhanced granularity is needed, capturing not just food but also seasonings, portion sizes, timing, eating speed, and associated GI symptoms or stress levels, particularly for conditions like IBS. Better data integration and visualization are crucial, moving beyond "eyeballing" trends to clear graphs showing changes in weight, calories, and nutrient ratios over time. D2 explained, \textit{"I eyeball the trends. It's helpful to see these numbers in a graphical form. Calorie differences and trends week by week—see how much they increased."} 

Dietitians emphasized the need for patient-centered, adaptive consultations that track motivation drivers and use self-tracking data for gradual changes. They also highlighted the importance of a mindful approach, such as using a \textit{"hunger and fullness scale (1-10 scale) tracking on a per-meal basis"} [D1] and looking at emotional hunger. D3 stated, \textit{"It would be nice to see a graph of energy as they were eating\dots We can compare it to normal eating habit graphs."} This shows a desire for tools that provide insights beyond basic caloric intake.

Furthermore, dietitians highlighted the importance of understanding the connection between food intake and the patient's subsequent emotional and physical state. There is a need for tools that allow patients to easily log not just what they ate, but also how they felt afterward—both emotionally (e.g., stressed, satisfied, guilty) and physically (e.g., energy levels, specific GI symptoms like bloating or pain). D1 noted that they look at the \textit{"emotional hunger and symptoms related to medication."} Capturing this information alongside dietary data could provide valuable insights into food triggers, sensitivities, and the complex relationship between diet, mood, and physical well-being.

\subsubsection{Food Journalers}
The interviews revealed that nutritional tracking is a complex and highly personal process for individuals driven by goals related to health and wellness. A key shared theme was the motivation to gain control and awareness over one's diet, often spurred by a desire to optimize physical performance or simply feel better. As L1 stated, \textit{"Once you start cooking for yourself, you have more control over how healthy you eat."} 

The data highlights a reliance on technology, with participants frequently using apps like MyFitnessPal and Lose It. However, this reliance is met with significant challenges, primarily related to inaccurate data and the tedious nature of manual logging. As L2 noted, \textit{"I don't think it's very accurate because the apps miscalculate on protein and calorie intake."} Participants struggle with the time-consuming process of inputting information, especially for home-cooked meals or when dining out, leading to tracking fatigue. L3 explained, \textit{"[I get deterred] When I’m hungry, it takes time and energy and thought. It’s sometimes tedious to find the exact product that I am eating."} Busy days and snacks are especially difficult to track accurately, and L1 pointed out that \textit{"Snacks are difficult to remember."}

A clear desire for more personalized and effortless tracking was evident, with suggestions for features like photo-based food analysis and integration with other devices. L3 wished to \textit{"upload it directly from my scale or if I could just take a pic and it could tell me."} While some apps offer photo features, journalers find them inaccurate, with L2 calling MyFitnessPal's feature a \textit{"scam kind of - it didn’t really know the accurate measurement."} The challenge of estimating portion sizes was also a common theme. L1 expressed frustration with the difficulty of determining food weights, noting, \textit{"You can’t bring a scale with you everywhere."}

Furthermore, the findings show a strong preference for visual feedback and actionable insights over raw numbers. L1 stated, \textit{"Numbers are good, but visualizations are easier to understand."} Participants also desire personalized progress indicators, such as \textit{"Progress pictures or indicators on the app that I’ve reached my goal (protein goal, fiber goal, calorie goal, etc.)"} [L4]. 

The ultimate goal for many was not just to log data, but to feel a sense of mental ease and accomplishment without the obsession that can sometimes accompany meticulous tracking. L2 explained, \textit{"If I thinks I'm feeling obsessive and compulsive... I'll take a break. It's bad to focus too much on the numbers."} L5 echoed this sentiment, stating, \textit{"Nutritional tracking can be very harmful, I would want to take away the obsessive manner of tracking."}

\subsection{Implications for Design of Self-Tracking Tools}
The findings from our interviews with both dietitians and food journalers reveal a clear set of design implications for future dietary self-tracking tools. 
The current landscape of tools, both manual and digital, fails to meet the core needs of accuracy, ease of use, and a holistic approach to health.
They lack the flexibility to adapt to a user's changing needs and the context awareness to understand the "why" behind the "what," resulting in a rigid, high-friction experience.

\subsubsection{Effortless and Accurate Data Capture}
The most significant barrier to consistent tracking is the high cognitive and physical effort required. 
Both dietitians and journalers highlighted this, with D2's advice to \textit{"Think about the least amount of work patients need to do"} and L3's complaint that when hungry, tracking \textit{"takes time and energy and thought."}
Future tools must tackle this by leveraging AI to provide a flexible logging process.
Language models for multimodal interaction allow users to log meals conversationally through text or audio, offering input flexibility and eliminating the tedious and often inaccurate process of manual entry.
As L2 noted, current photo features are a \textit{"scam,"} and L1 complained about the difficulty of weighing food. By using a pre-trained model to semantically search a robust database of food images and their nutritional data, models can provides a more reliable and accurate way for users to \textit{"just take a pic and it could tell me"} [L3], moving beyond simple visuals to provide a precise nutritional breakdown without the need for a scale.

\subsubsection{Actionable Visualizations and Personalized Feedback}
Simply logging data is not enough; users seek meaningful insights that connect their habits to their goals.
L1's desire to \textit{"check if I ate too much of a certain food"} and L4's preference for \textit{"Progress pictures or indicators on the app that I’ve reached my goal"} highlight a strong need for data visualization.
Dietitians echoed this, with D2 stating, \textit{"I eyeball the trends. It's helpful to see these numbers in a graphical form."}
By storing structured data on calories, protein, fat, and other metrics, systems can generate clear, intuitive graphs that show trends over time.
Systems can move from simple loggers to providing the type of visual feedback that adapts to the motivational context of a user and helps them feel a sense of mental ease and accomplishment.

\subsubsection{Integration of  Holistic, Contextual Data}

The interviews consistently revealed that food intake is just one part of the health equation. 
Dietitians emphasized the importance of tracking a wider range of contextual data, from emotional states to physical symptoms.
D1's focus on a \textit{"hunger and fullness scale"} and "emotional hunger" demonstrates the value of capturing this qualitative data.
A reliance on conversational AI interfaces can capture this holistic information.
Systems can be designed to ask about a user's emotional state or energy levels in the natural flow of conversation.
This comprehensive approach, combined with the ability to store a user's goals, allows for highly personalized prompts that directly address a user’s unique health motivations.

\subsubsection{Focus on a Positive and Non-Obsessive Approach}

A critical finding, particularly from the journalers, is the risk of tracking leading to an \textit{"obsessive and compulsive"} mindset [L2]. L5's warning that \textit{"Nutritional tracking can be very harmful"} is a powerful design constraint. By using conversational, non-numeric approaches, systems can reduce the focus on meticulous, number-driven logging, which can be a source of anxiety. Additionally, adding personalization to the prompts of these conversations allows systems to tailor its feedback to focus on a user's broader goals, such as feeling \textit{"in control"} [L1] or simply building a healthy lifestyle, rather than being a \textit{"slave to the app"} [L2]. By leveraging user goals and sentimental tracking [L5], the system can promote a healthy relationship with food, shifting the focus from perfect data to sustainable, positive habits.

\section{System Design}
\begin{figure*}[ht!]
    \centering
    \vskip -0.1in
    \includegraphics[width=1\linewidth]{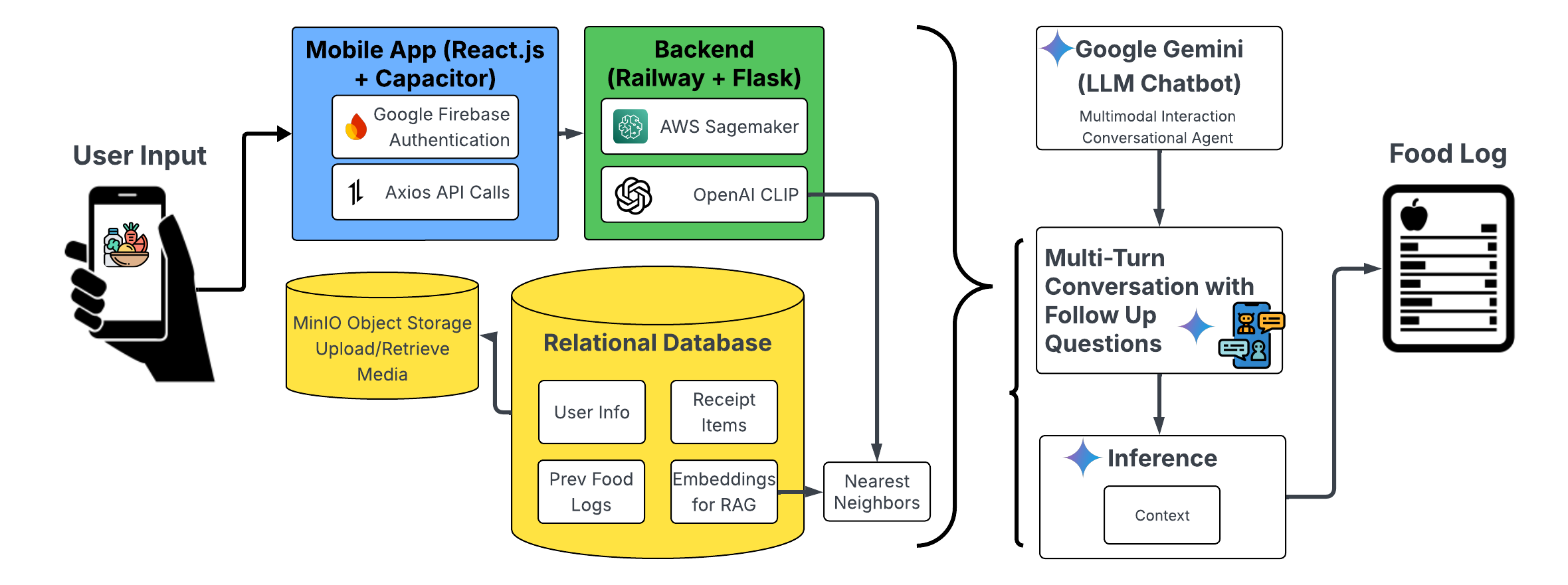}
    \caption{System overview: Users input multimodal food logs which are processed along with relevant context to extract nutritional information. This data is fed into an LLM to validate the nutritional information and determine if there is any missing information. Finally, all the information is sent to Gemini to generate food log data. The resulting data is displayed as individual food logs that can be examined and aggregate graphical visualizations.}
    \label{fig:system_flow}
    \Description[System overview: Users input multimodal food logs which are processed by an LLM.]{System overview: Users input multimodal food logs which are processed along with relevant context to extract nutritional information. This data is fed into an LLM to validate the nutritional information and determine if there is any missing information. Finally, all the information is sent to Gemini to generate food log data. The resulting data is displayed as individual food logs that can be examined and aggregate graphical visualizations.}
\end{figure*}

The SnappyMeal system employs a microservices-oriented architecture that separates the frontend, backend, and specialized AI services into modular components. This structure allows for independent development, deployment, and evaluation of major system components. An overview of key system modules and how they interact can be found in Figure~\ref{fig:system_flow}.

\subsection{Mobile App}

\begin{figure*}[ht!]
\centering
% (a) Trend Screen
\begin{subfigure}{.24\textwidth}
    \centering
    \includegraphics[width=\linewidth]{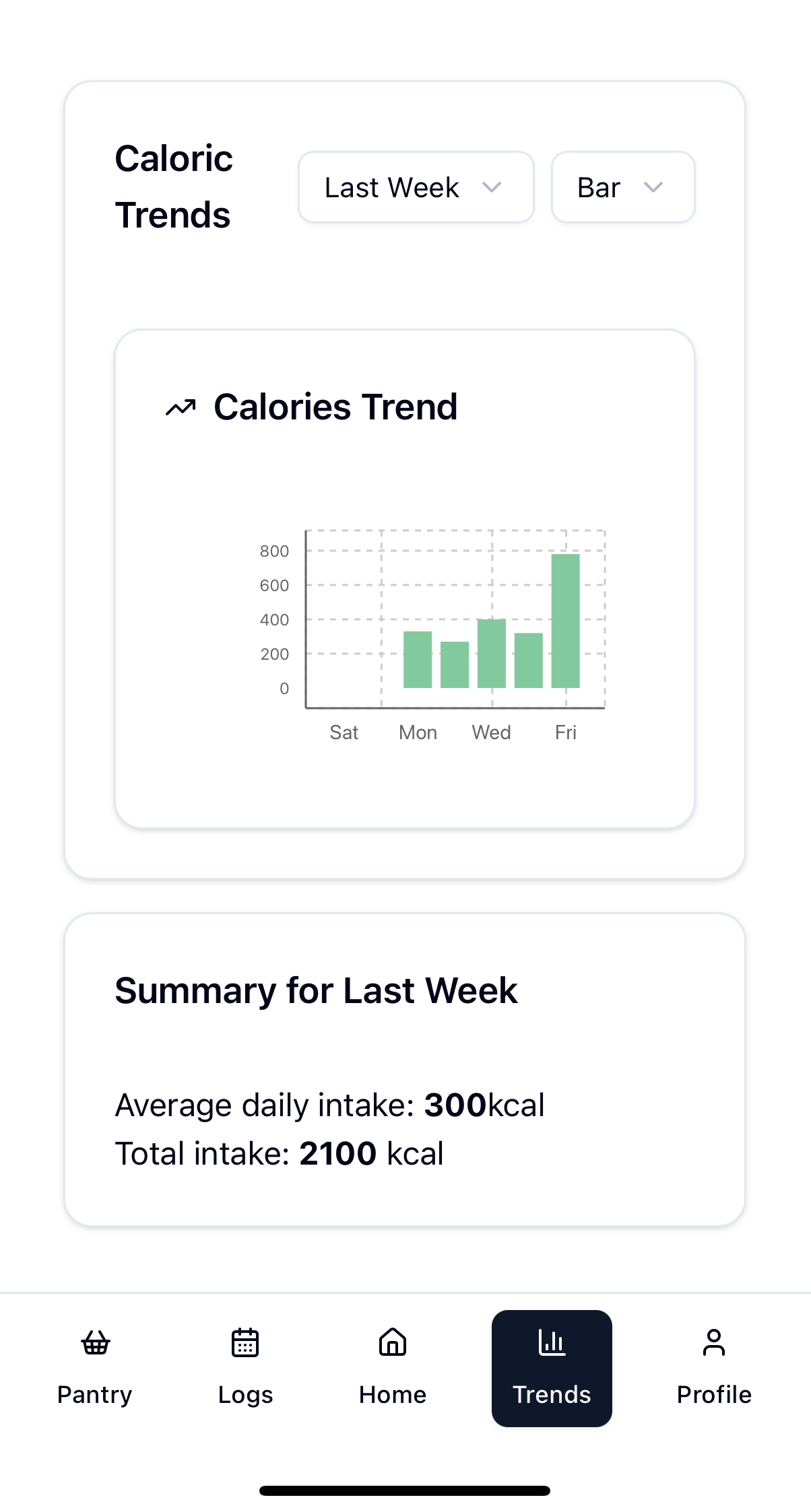}
    \caption{Trend Screen}
    \label{fig:trends_screen}
\end{subfigure}
\hfill % Adds horizontal space
% (b) Dashboard Screen
\begin{subfigure}{.24\textwidth}
    \centering
    \includegraphics[width=\linewidth]{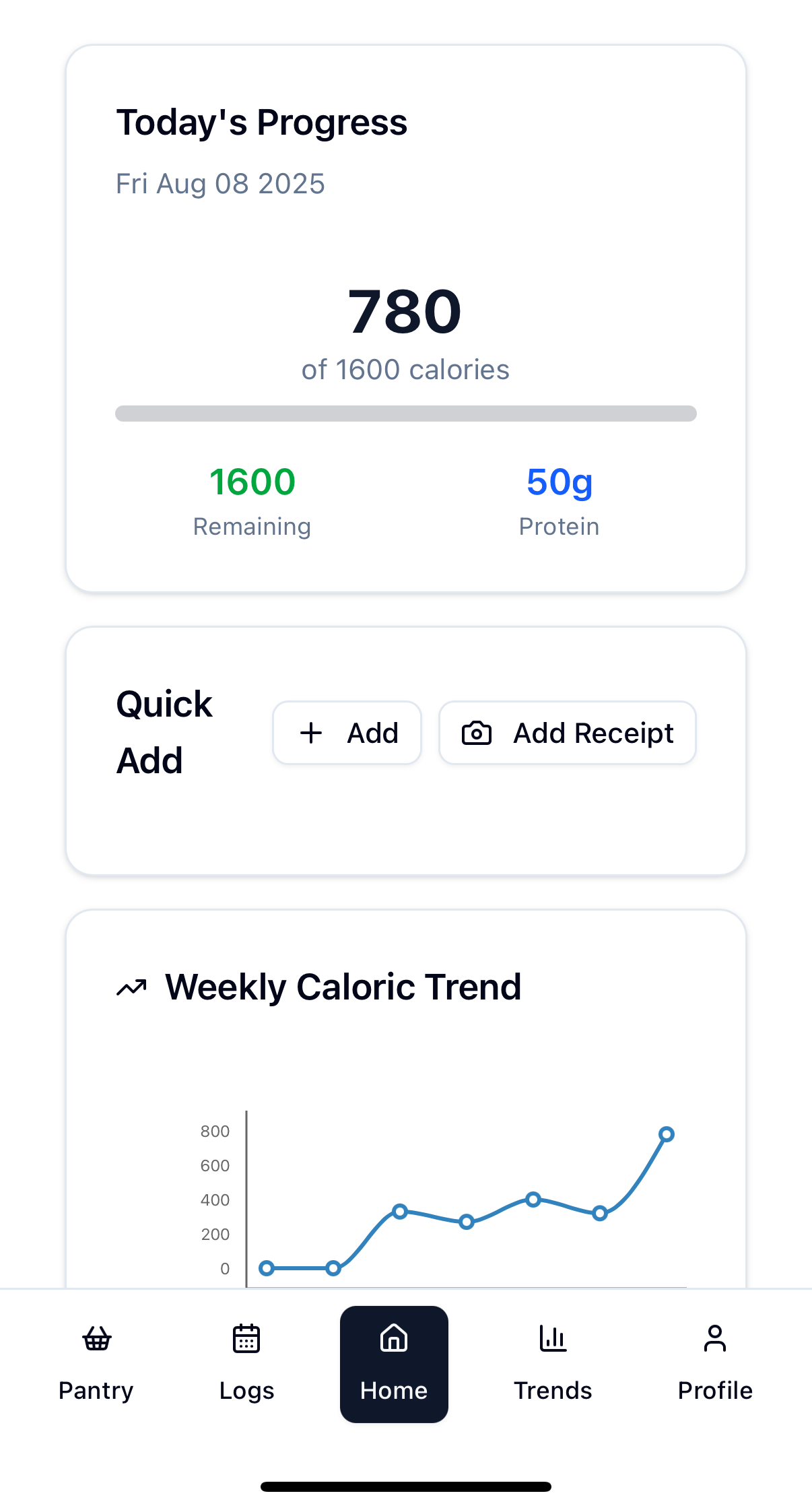}
    \caption{Dashboard Screen}
    \label{fig:dashboard_screen}
\end{subfigure}
\hfill % Adds horizontal space
% (c) Logging Screen
\begin{subfigure}{.24\textwidth}
    \centering
    \includegraphics[width=\linewidth]{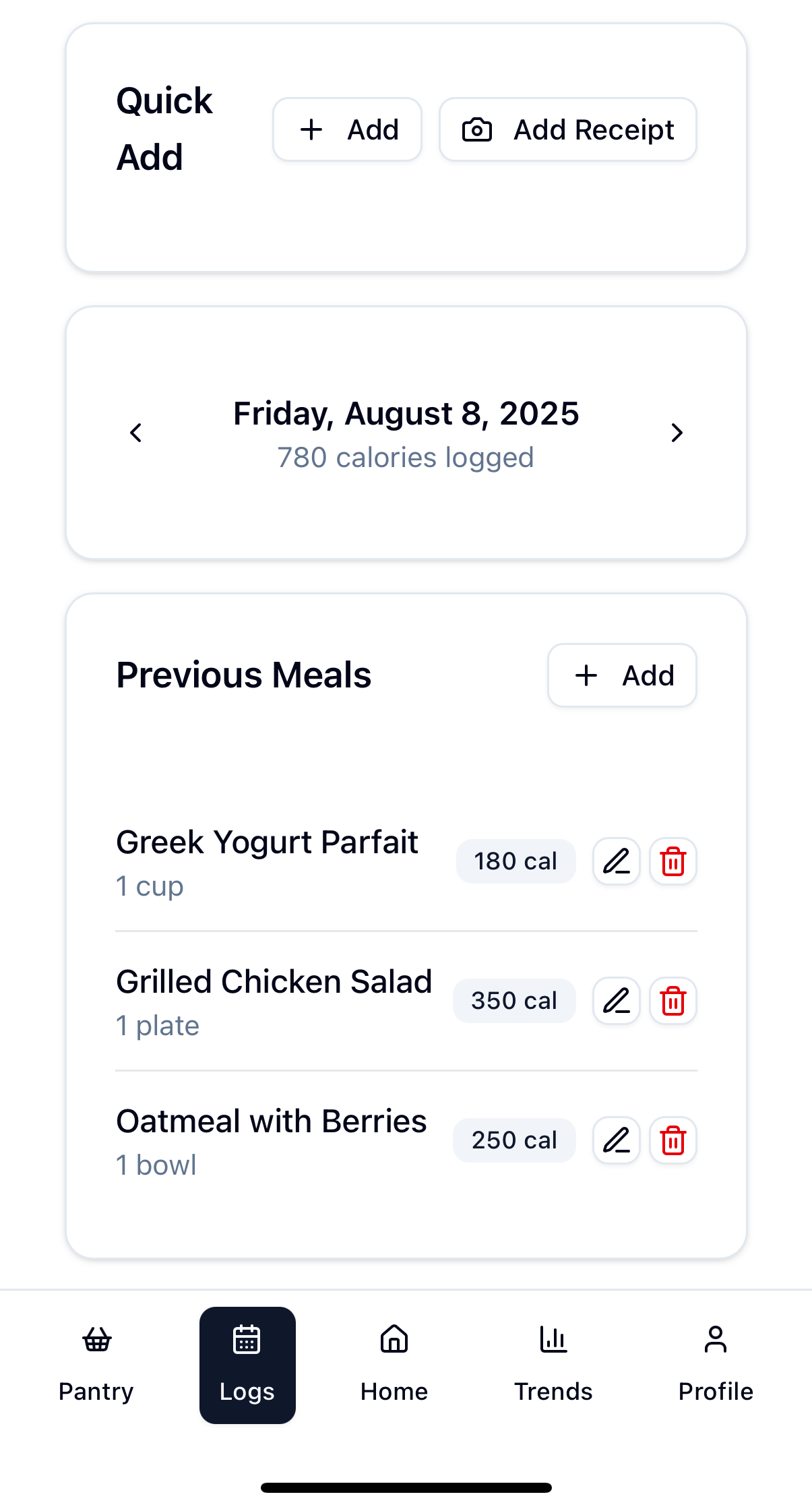}
    \caption{Logging Screen}
    \label{fig:logging_screen}
\end{subfigure}
\hfill % Adds horizontal space
% (d) Profile Screen
\begin{subfigure}{.24\textwidth}
    \centering
    \includegraphics[width=\linewidth]{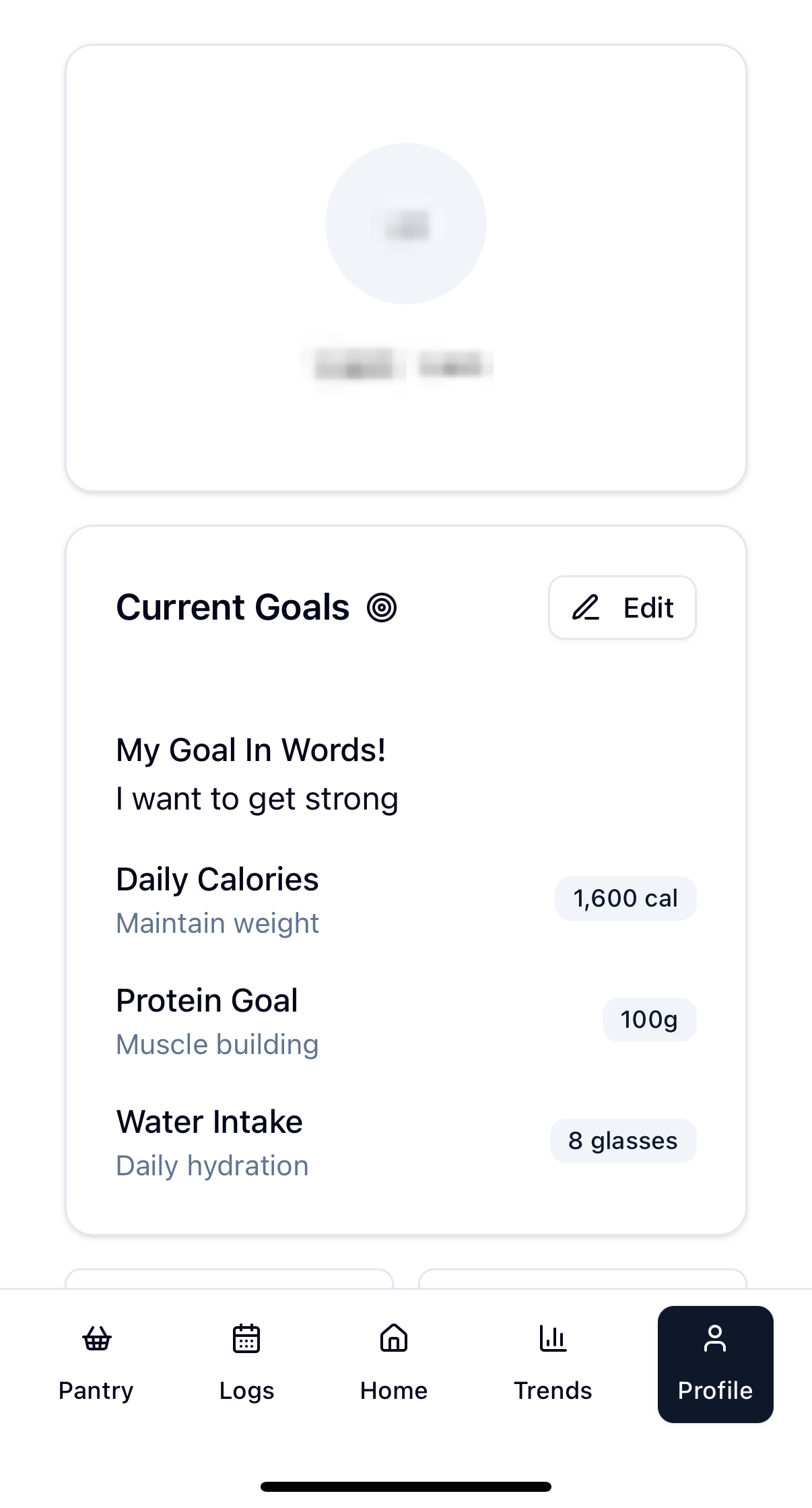}
    \caption{Profile Screen}
    \label{fig:profile_screen}
\end{subfigure}

% Add a main caption for the entire figure
\caption{Overview of the application's main interface screens.}
\label{fig:all_screens} 
\Description[Screenshots of the trends, dashboard, logging, and profile screens of the SnappyMeal app.]{Screenshots of the trends, dashboard, logging, and profile screens of the SnappyMeal app.}
\end{figure*}

We design our app with 5 screens: a pantry screen where users can view their uploaded receipt items, a log (\ref{fig:logging_screen}) screen where users can view their generated food logs, a dashboard screen (\ref{fig:dashboard_screen}) where users can see general progress charts as well as efficiently add new food logs or receipt uploads, a trend screen (\ref{fig:trends_screen}) where users can see trends and visualizations about their nutritional data, and a profile screen (\ref{fig:profile_screen}) where users can view and edit their personal information and goals. 
When users first create an account, they input their numeric and personal nutrition goals. 
This information forms the initial context which is used to tailor prompts, ask relevant follow-up questions, and provide support to help a user achieve these goals. 
When users log food, they are offered input flexibility, allowing them to upload an image, text description, or audio description of their meal. 
Subsequently, an LLM (Gemini), examines the uploaded media to determine whether enough information is present. 
If not, the model generates a follow up question to gain a better understanding of the media. Finally, the original media, user's nutritional goals, receipt context, and clarifying conversation history are sent to Gemini to generate a comprehensive nutrition log.

% \textcolor{red}{VI: TODO come back and fix this}
% We create a mobile application for users to easily input multimodal data about their 
% When the user first downloads the app, they are directed to create an account\ref{authentication}, but also provide personal information relevant to nutrition tracking such as their height, weight, age, calorie goals, protein goals, water goals, and a quick blurb describing their general dietary goals.

% This mobile application is built using React JS and is encapsulated with Capacitor to enable cross-platform deployment on both iOS and Android. This choice ensures a consistent user experience while leveraging web-based development efficiency. Axios is used for asynchronous API calls to the backend, handling user interactions such as logging meals, viewing data, and uploading images. Screenshots from the app interface can be found in Figure \ref{fig:all_screens}.

%\subsection{Backend} The backend is developed in Python using the Flask framework, providing a flexible foundation for the application's core logic. It is hosted on a Railway server, which simplifies deployment and management. The backend is responsible for user authentication, processing user requests, interacting with the database, and orchestrating calls to the AI models.

%\subsection{Authentication} \label{authentication} User registration and authentication are handled by Google Firebase Authentication. This choice offloads the complexity of managing user credentials, password hashing, and secure session management. The frontend interacts directly with Firebase to handle user signup and login. Once authenticated, a user's unique Firebase ID token is sent to the backend with each request. The backend then verifies this token using the Firebase Admin SDK to ensure the user is authenticated and authorized to access their data.

\subsection{Data Storage} A relational database is used to store structured user data, meal logs, and conversation history. This choice provides strong data integrity and transactional consistency. The provided schema (see Appendix~\ref{appendix:food_logging_relation_schema}) outlines the relationships between key entities like \textit{users}, \textit{food\_logs}, and \textit{receipt\_items}.
\begin{itemize}
    \item \textbf{\textit{users}:} Stores core user information, including demographics (\textit{age}, \textit{height}, \textit{weight}), health goals (\textit{target\_calories}, \textit{target\_protein}, \textit{target\_water}), and a free-form \textit{text\_goals} field for personalized prompts. The \textit{user\_id} serves as the primary key.

\item \textbf{\textit{food\_logs}:} This is the central table for meal tracking. Each entry records a single meal, including the nutritional breakdown (\textit{calories}, \textit{protein}, \textit{fat}, etc.), the meal's name, and a link to the \textit{chat\_history} and any associated \textit{media\_url} for multimodal logging.

\item \textbf{\textit{receipt\_items}:} This table stores data extracted from user-uploaded grocery receipts. It helps the system understand the user's general dietary habits and food preferences by linking items purchased at a store to a user's ID.

\item \textbf{\textit{conversations}:} This table stores the history of user interactions with the Gemini API. It records each conversational turn, including the original prompt and the API's response, allowing for context-aware, follow-up interactions.

\item \textbf{\textit{food\_embeddings}:} This is a crucial table for the RAG functionality. It stores pre-computed vector embeddings for a large corpus of food items. These embeddings, generated by the CLIP model from the Nutrition5k\cite{thames2021nutrition} dataset, allow the system to efficiently find the most semantically similar food items based on a user-uploaded image embedding. The table also includes metadata such as \textit{food\_label}, \textit{estimated\_calories}, and other nutritional data sourced directly from the dataset.

\item \textbf{\textit{personalized\_prompts}:} This table stores the user's personalized prompt derived from their goals during signup. These prompts are subject to change if user goals are editted, providing flexibility and ensuring the system's context awareness is up-to-date.
\end{itemize}

\noindent A separate object storage system, MinIO, is used to store unstructured data such as user-uploaded images and audio files. This offloads large binary data from the primary database, improving database performance and allowing for efficient retrieval and processing by the AI models.

\subsection{Multimodal Logging} 

To achieve effortless, accurate, and contextually-aware data capture—the primary goals identified in our formative study—the SnappyMeal system is designed to accept and process food logs across multiple modalities: image, text, and audio. Our approach fundamentally relies on multimodality, which allows the system to process and relate different data types.

\subsubsection{Image inputs}
The OpenAI CLIP (Contrastive Language–Image Pre-training)\cite{radford2021learning} model, hosted on AWS Sagemaker, is used for Retrieval-Augmented Generation (RAG). CLIP's fundamental contribution to downstream applications stems from its ability to map text and images into a shared embedding space where their vector representations are aligned based on semantic content. This allows us to use the same model for text and image modalities.  When a user uploads an image, the CLIP model generates an image embedding. 
This embedding is then used to perform a semantic search using cosine similarity against the \textit{food\_embeddings} table to find the most relevant food items in the database. For this purpose, we used the Nutrition5k\cite{thames2021nutrition} dataset, a large-scale, pre-annotated dataset of food images and their corresponding nutritional information, to create the foundational \textit{food\_embeddings} for our database. This dataset serves as the knowledge base for the RAG system, significantly improving the accuracy of food identification. Using a nearest neighbors search, the system identifies the most semantically similar foods in the database and presents its verified nutritional information as context in the aggregate prompt. 

\subsubsection{Text inputs}
Similarly, when a user uploads a text description of their meal, the CLIP model generates an embedding. Due to CLIPS contrastive nature and its training on text-image pairs, we are able to match text embeddings to image embeddings. Again, using a nearest neighbors search, the system identifies foods in the database that are most semantically similar to the users' input  and presents their verified nutritional information as context in the aggregate prompt. Unlike with images, there are no food-text embeddings in the vector database, so we rely on the multimodality of CLIP.

\subsubsection{Audio inputs}
When a user uploads an audio description, audio file is directly sent to the aggregate Gemini prompt input with the user context since CLIP does not support audio inputs. The prompt includes instructions explaining that this file is user uploaded media. While we could have included speech-to-text preprocessing to utilize CLIP, the additional step would have increased latency making it challenging to provide users real-time interaction. 

\subsection{Interactive prompting}
The Gemini API is the core of the app's multimodal interaction and personalized prompting. It processes natural language queries (text and audio) and analyzes uploaded images to identify food items and provide conversational feedback. For this task, we use Gemini 2.5-Flash due to its balance of strong multi-modal support. While Gemini 2.5-Pro is more powerful for many tasks, it would have introduced more latency and cost.

\subsubsection{Follow-up Questions}
\label{follow_up_q_section}
By default, every new food log receives follow-up question with the intention of clarifying any missing information from the raw uploaded media. This process occurs through a Gemini multi-turn conversation stemming from the prompt in Appendix~\ref{appendix:follow_up_question_generation_prompt}. Contrary to single-turn LLM calls, the multi-turn conversation has a sense of conversation history that allows the model to lead goal-oriented "conversations."  After every answer, the model is asked if it has enough information to generate a comprehensive food log; if not, it is asked to generate another follow up question. This conversation history is then included in the aggregate prompt to provide the model with additional context and improve the accuracy.

\subsubsection{Receipt Context}
Users are asked to upload their receipts into the system to improve their generated food logs. Receipts are parsed using Gemini to extract name, quantity, and source (where they were bought). Utilizing the publicly available nutrition information of many popular grocery store products, we use Gemini to generate nutritional summaries of each of the purchased food ingredients. These ingredient data are then added as context in the aggregate prompt to improve accuracy. Sometimes the follow-up questions can clarify if the meal being analyzed was cooked using any pantry ingredients.

\section{System Evaluation}
\label{section:system_eval}

To assess the performance and reliability of the nutrition tracking software, we carried out a systematic technical evaluation focused on quantifying its accuracy in food recognition and nutritional estimation. This assessment used the publicly available Nutrition5k dataset, a benchmark repository comprising 3,490 food images meticulously paired with ground-truth nutritional information. 
The following section details the specific experimental setup, the definition of the performance metrics employed, and the results derived from challenging the software against this diverse labeled data corpus.

\subsection{Experiment Setup}

The technical evaluation was conducted utilizing the Gemini Batch API to facilitate efficient inference across the entire Nutrition5k dataset. A standardized, zero-shot prompt was designed to instruct the model to identify food items and output the corresponding nutritional breakdown in a structured JSON format.

To determine the impact of individual architectural components on overall performance, we performed a controlled ablation study. This involved comparing the baseline model's performance against configurations where key features were selectively introduced:

\subsubsection{Baseline Model:} To test the core visual recognition and vanilla LLM inference, we sent in the same prompt (see Appendix~\ref{appendix:vanilla_prompt}) for each food image instance.
    
\subsubsection{Ingredient Addition: } The model was augmented with the capacity to infer and explicitly list constituent ingredients beyond the main food item, such as those present in a receipt. The Nutrition5k dataset provides ground-truth ingredients for each image. To mitigate bias toward the visible ingredients, we deliberately introduced negative sampling. Specifically, we began by establishing a repository containing all unique ingredients present across the dataset, denoted as $\{Ing\}_{\text{all}}$. For an image $i$ containing $k$ true ingredients ($|\{Ing\}_{i}|=k$), an equivalent set of $k$ negative samples was randomly drawn from the set of all ingredients not present in the image, $\{Ing\}_{\text{all}} - \{Ing\}_{i}$. The ingredients and their nutritional values were included in the original prompt.

\subsubsection{Retrieval-Augmented Generation (RAG):} Since we had used the Nutrition5k dataset as our vector database, we performed RAG by omitting the image being evaluated and finding the nearest images based on OpenAI's clip model. When evaluating image $i$, we performed cosine similarity nearest neighbor search on images indexed $0, 1, \dots, i-1, i+1, \dots, n$. The top 5 closest matches and their nutritional information were added into the prompt as additional context for the model.

\subsubsection{Follow Up Questions:} Due to the large size of the dataset, we could not answer follow-up questions for every image. Instead, we randomly sampled 100 images. Four members of the research team split these 100 samples, and manually answered the generated follow-up questions. This ensured a diversity in the answers of the questions to further generalize our evaluation. Upon completing this, the vanilla prompt was sent with the answered question for evaluation in the following format.

\noindent\texttt{Here is a clarifying question and answer that can help you better understand the food:\\\{question\}\\\{answer\}}

\subsection{Peformance Metrics} Nutritional performance was quantified using the Mean Absolute Error (MAE) between the estimated calories (kcal), protein (g), fat (g), and carbohydrates(g) generated by our experiments and the ground-truth values from the Nutrition5k dataset. We also evaluate the Root Mean Squared Error (RMSE) which is more sensitive to outliers (see Appendix \ref{appendix:equations-error} for~equations).

To assess the statistical significance of our results, we constructed 95\% confidence intervals using the percentile bootstrap method. This technique allows us to estimate the uncertainty of the MAE and RMSE given the varying size of our evaluation sets. The procedure involved generating $B$ bootstrap samples (in our case $B = 1000$) by sampling with replacement from the original $n$ data points. For each of these $B$ samples, we re-calculated our metric, which yielded a distribution of $B$ metric estimates.
The 95\% confidence interval was then derived directly from this distribution by taking the 2.5th and 97.5th percentiles as the lower and upper bounds, respectively. A formal description of this method is provided in Appendix~\ref{appendix:equations-bootstrap}.

\subsection{Results}
\begin{table}[ht!]
\caption{Evaluation Metrics and Confidence Intervals of Individual Model Performance by Nutritional Value  $n=3466$.}
\label{table:ci_results}
\begin{tabular}{llrr}
\toprule
 Nutritional Value & Model & MAE (95\% CI) & RMSE (95\% CI) \\

\midrule
\multirow[t]{4}{*}{Calories (kcal)} & vanilla & 120.38 (115.54, 124.92) & \textbf{188.58} (171.16, 213.45) \\
 & receipt & 121.57 (116.47, 127.20) & 192.87 (175.24, 220.50) \\
 & RAG & \textbf{120.21} (115.40, 125.23) & 188.94 (171.73, 214.79) \\
\cline{1-4}
\multirow[t]{4}{*}{Protein (g)} & vanilla & 7.72 (7.39, 8.10) & 12.86 (12.17, 13.50) \\
 & receipt & 7.67 (7.32, 8.03) & 12.73 (12.11, 13.32) \\
 & RAG & \textbf{7.40} (7.05, 7.75) & \textbf{12.32} (11.67, 12.97) \\
\cline{1-4}
\multirow[t]{4}{*}{Carbohydrates (g)} & vanilla & 12.33 (11.69, 13.04) & 23.75 (18.10, 31.42) \\
 & receipt & 12.46 (11.79, 13.20) & 24.21 (18.68, 32.18) \\
 & RAG & \textbf{12.19} (11.56, 12.97) & \textbf{23.48} (18.00, 31.34) \\
\cline{1-4}
\multirow[t]{4}{*}{Fat (g)} & vanilla & 7.98 (7.66, 8.27) & 12.36 (11.82, 12.83) \\
 & receipt & 8.05 (7.72, 8.38) & 12.50 (11.94, 13.07) \\
 & RAG & \textbf{7.81} (7.50, 8.13) & \textbf{11.90} (11.43, 12.38) \\
\cline{1-4}
\bottomrule
\end{tabular}
\end{table}

\begin{table}[ht!]
\caption{Ablation Evaluation Metrics and Confidence Intervals of Model Performance by Nutritional Value ($n=100$).}
\label{table:ci_results_n_100}
\begin{tabular}{llrr}
\toprule
 Nutritional Value & Model & MAE (95\% CI) & RMSE (95\% CI) \\

\midrule
\multirow[t]{4}{*}{Calories (kcal)} 

 & vanilla & 148.88 (122.51, 177.29) & 208.00 (173.08, 243.00) \\
 & receipt & 148.42 (117.40, 180.84) & 219.67 (176.78, 260.31) \\
 & RAG & 145.45 (114.11, 180.28) & 223.14 (173.50, 270.91) \\
 & follow-up & 161.82 (123.14, 209.24) & 274.48 (197.67, 358.78) \\
 & RAG + follow-up & 168.87 (128.05, 218.25) & 288.50 (199.23, 377.33) \\
 & receipt + follow-up & 144.79 (109.99, 187.02) & 242.07 (174.63, 323.50) \\
 & RAG + receipt & \textbf{123.96} (97.16, 153.03) & \textbf{191.92} (149.13, 232.17) \\
 & RAG + receipt + follow-up & 153.00 (111.04, 199.29) & 273.19 (181.77, 359.27) \\
\cline{1-4}
\multirow[t]{4}{*}{Protein (g)}
 & vanilla & 10.06 (7.92, 12.22) & 14.72 (11.88, 17.39) \\
 & receipt & 9.38 (7.37, 11.84) & 14.72 (11.57, 17.96) \\
 & RAG & 9.05 (7.06, 11.11) & 13.98 (10.77, 16.99) \\ 
 & follow-up & 10.27 (8.03, 12.57) & 15.64 (12.01, 18.94) \\
 & RAG + follow-up & 10.60 (8.24, 13.23) & 16.92 (13.08, 20.59) \\
 & receipt + follow-up & 9.02 (7.01, 11.12) & 14.10 (10.58, 17.22) \\
 & RAG + receipt & \textbf{8.75} (6.87, 10.88) & \textbf{13.66} (10.77, 16.57) \\
 & RAG + receipt + follow-up & 9.08 (6.82, 11.60) & 15.06 (11.16, 18.64) \\
\cline{1-4}
\multirow[t]{4}{*}{Carbohydrates (g)} 
 & vanilla & 15.30 (12.44, 18.76) & 22.77 (18.17, 27.82) \\
 & receipt & 15.81 (12.50, 19.27) & 23.76 (18.29, 29.62) \\
 & RAG & \textbf{13.80} (10.43, 17.29) & \textbf{22.06} (16.64, 26.81) \\
 & follow-up & 18.00 (13.31, 23.49) & 31.99 (20.92, 43.00) \\
 & RAG + follow-up & 17.21 (12.56, 22.87) & 31.75 (20.58, 42.51) \\
 & receipt + follow-up & 16.43 (12.49, 21.42) & 28.27 (19.53, 37.60) \\
 & RAG + receipt & 13.81 (10.36, 17.53) & 22.43 (16.63, 27.92) \\
 & RAG + receipt + follow-up & 16.56 (11.79, 21.72) & 31.00 (19.40, 40.69) \\
\cline{1-4}
\multirow[t]{4}{*}{Fat (g)} 
 & vanilla & 8.05 (6.53, 9.86) & 11.73 (9.74, 13.76) \\
 & receipt & 8.56 (6.59, 10.53) & 12.95 (10.23, 15.50) \\
 & RAG & 8.88 (6.89, 10.93) & 13.61 (10.62, 16.37) \\
 & follow-up & 9.58 (7.35, 12.15) & 15.31 (11.93, 18.93) \\
 & RAG + follow-up & 9.42 (7.20, 11.91) & 15.26 (12.13, 18.38) \\
 & receipt + follow-up & 8.23 (6.25, 10.20) & 13.25 (9.94, 16.19) \\
 & RAG + receipt & \textbf{7.17} (5.50, 8.98) & \textbf{11.34} (8.59, 14.10) \\
 & RAG + receipt + follow-up & 8.45 (6.24, 10.77) & 14.63 (10.84, 18.00) \\
\cline{1-4}
\bottomrule
\end{tabular}
\end{table}

The evaluation results of the full dataset can be found in Table~\ref{table:ci_results} where the point metrics and their confidence intervals are reported. 
Due to follow-up involving human input, we only evaluated the feature on 100 samples.
The results of the ablation study can be found in Table~\ref{table:ci_results_n_100} where we evaluated every feature and some combinations on the same 100 food image samples. We highlight some specific examples where follow-up questions help and hurt in Table~\ref{table:food_analysis_alt}.

\subsection{Discussion}
\label{section:system_eval_discussion}
\begin{table}[htbp]
\centering
\caption{Food Image Samples from the Nutrition5k Dataset where follow-up questions caused benefit or friction. In the $\Delta$ MAE column, a \textcolor{green}{$\downarrow$} means improvement (MAE decreased), a \textcolor{red}{$\uparrow$} means decline (MAE increased), and a $\bigcirc$ means no change.}
\label{table:food_analysis_alt}

\begin{tabular}{ c c c c c c} 
 \toprule

 Food Image & Follow-up Q\&A & Metric & vanilla MAE & follow-up MAE & MAE Improvement\\
\midrule
\\
 
    % Used \multirow for the first 2 columns
\multirow{2}{*}{\includegraphics[width=0.2\linewidth]{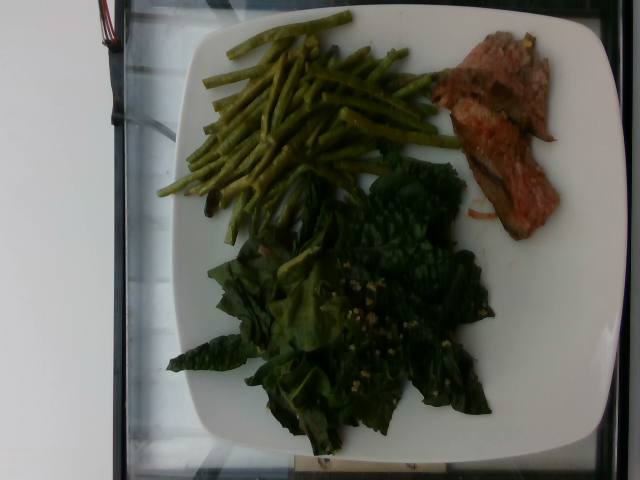}} & 
\multirow{2}{*}{\shortstack[l]{SnappyMeal: \textit{What kind} \\ \textit{of meat is that?}\\ User: \textit{Beef.}}} & 
 
Calories (kcal) & 240.46 & 112.46 & \textcolor{green}{$\downarrow$} \\

  & & Protein (g)  & 19.46 & 17.56 & \textcolor{green}{$\downarrow$} \\ 
  & & Carbohydrates (g) & 6.20 & 2.90 & \textcolor{green}{$\downarrow$} \\ 
  & & Fat (g) & 16.07 & 4.27  & \textcolor{green}{$\downarrow$} \\ 
  \\
  \\
  \\
  \cline{1-6}
  \\
  \multirow{2}{*}{\includegraphics[width=0.2\linewidth]{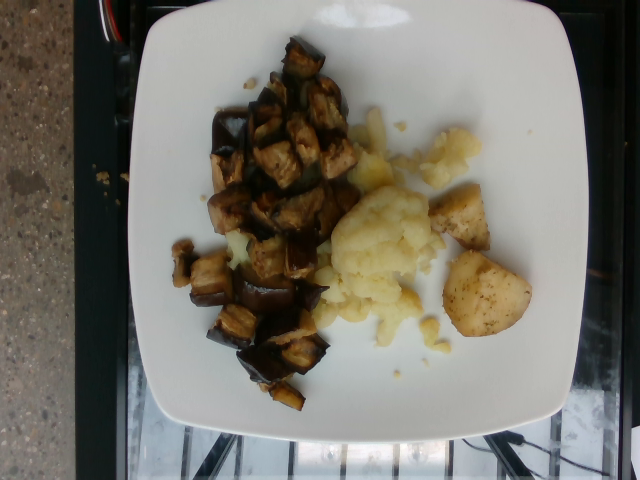}} & 
\multirow{2}{*}{\shortstack[l]{SnappyMeal: \textit{How were} \\ \textit{the vegetables prepared?}\\ User: \textit{Fried eggplants and} \\ \textit{steamed cauliflower.}}} & 
 
Calories (kcal)  & 389.83 & 234.83 & \textcolor{green}{$\downarrow$} \\

  & & Protein (g)  & 6.57 & 0.57 & \textcolor{green}{$\downarrow$} \\ 
  & & Carbohydrates (g)  & 58.88 & 20.88 & \textcolor{green}{$\downarrow$} \\ 
  & & Fat (g) & 17.45 & 14.45 & \textcolor{green}{$\downarrow$} \\ 
  \\
  \\
  \\
  \cline{1-6}
  \\
  \multirow{2}{*}{\includegraphics[width=0.2\linewidth]{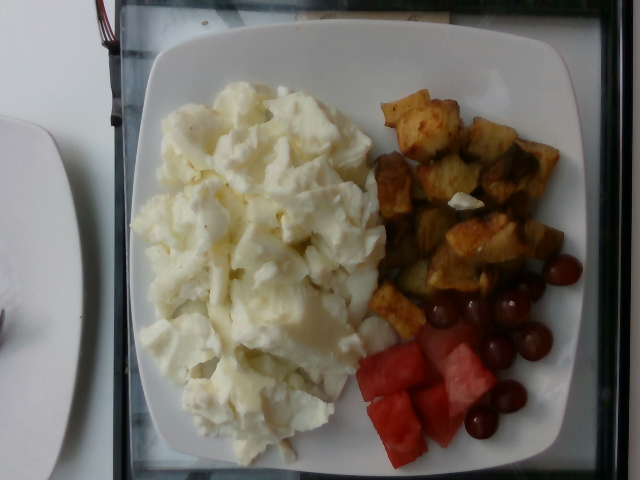}} & 
\multirow{2}{*}{\shortstack[l]{SnappyMeal: \textit{Are these} \\ \textit{egg whites or whole eggs?}\\ User: \textit{Egg whites.}}} & 
 
Calories (kcal) & 45.81 & 138.89 & \textcolor{red}{$\uparrow$} \\

  & & Protein (g) & 24.25 & 25.25 & \textcolor{red}{$\uparrow$} \\ 
  & & Carbohydrates (g) & 7.96 & 7.96 & $\bigcirc$ \\ 
  & & Fat (g) & 4.79 & 0.21 & \textcolor{green}{$\downarrow$} \\ 
  \\
  \\
  \\
  \cline{1-6}
  \\
  \multirow{2}{*}{\includegraphics[width=0.2\linewidth]{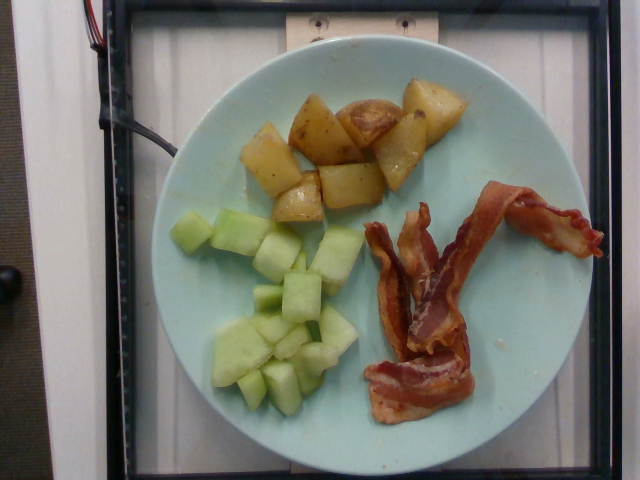}} & 
\multirow{2}{*}{\shortstack[l]{SnappyMeal: \textit{How many} \\\textit{strips of bacon did you eat?}\\ User: \textit{Three.}}} & 
 
Calories (kcal)  & 101.06 & 173.06 & \textcolor{red}{$\uparrow$} \\

  & & Protein (g)  & 2.09 & 5.09 & \textcolor{red}{$\uparrow$} \\ 
  & & Carbohydrates (g)  & 18.046 & 35.046 & \textcolor{red}{$\uparrow$} \\ 
  & & Fat (g)  & 2.89 & 6.89 & \textcolor{red}{$\uparrow$} \\ 
  \\
  \\
  \\
  \cline{1-6}
 
 \bottomrule
\end{tabular}
\Description[Food Image Samples from the Nutrition5k Dataset where follow-up questions caused benefit or friction.]{Food Image Samples from the Nutrition5k Dataset where follow-up questions caused benefit or friction. The first image depicts a plate of greens and beef. The second image depicts a plate of brussel eggplant, cauliflower, and potatoes. The third image depicts a plate of egg whites, potatoes, and carrots. The fourth plate is depicts a plate of melons, potatoes, and bacon.}
\end{table}

In summary, the RAG model performed the best for protein, carbohydrate, and fat estimations in Table~\ref{table:ci_results}.
However, the confidence intervals mostly overlapped, so we cannot confidently conclude that RAG is the best model.
Additionally, we noticed that the error values increased when evaluating the same model on the $n=3466$ dataset and the $n=100$ dataset.
This suggests the 100 images evaluated in the smaller dataset can be some of the more unclear images for the model.
In Table~\ref{table:ci_results_n_100}, the overlap for RAG and receipt is not as pronounced. 
We noticed it has the best performance of all models for each nutritional category except carbohydrates.
This is likely because those two features combined provide the most numeric nutritional information without introducing extra textual noise.
Unlike receipts alone, the RAG data in "RAG and receipts" helps standardize the input by providing a grounding estimation.

Contrary to expectations, we observed no conclusive evidence that the follow-up questions directly improved nutrition estimation. This outcome is likely tied to two major factors: the model's small evaluation sample size and the observed behavior of the LLM.
The confidence interval analysis suggested that the model may be second-guessing itself; generating a question, receiving an imperfect or ambiguous answer from the user, and then allowing that conflicting information to degrade the final estimation rather than refine it. 
A clear benefit of the follow-up questions is observed in the first two examples (rows 1 and 2) of Table~\ref{table:food_analysis_alt}, where the follow-up question successfully disambiguated key nutritional factors that are difficult or impossible to determine from visual data alone.
In the first case, the baseline model's estimation for a meat dish was significantly improved after the user clarified the food type as "Beef." The initial "vanilla MAE" was high, suggesting the model may have defaulted to a generic "meat" profile or an incorrect specific type (e.g., chicken). The user's textual input allowed the model to apply a more accurate nutritional profile, resulting in a uniform improvement across all four measured metrics (Calories, Protein, Carbohydrates, and Fat).
Similarly, the query "How were the vegetables prepared?" provided critical, non-visual context. The user's response, "Fried eggplants and steamed cauliflower," resolved ambiguity about preparation methods that significantly impact nutritional content. The baseline model cannot visually distinguish "steamed" from "fried," "boiled," or "roasted." The clarification allowed for a major correction, particularly in fat and calorie estimations, and again resulted in a uniform improvement.

Conversely, the final two examples (rows 3 and 4) of Table~\ref{table:food_analysis_alt} illustrate moments where the model's performance degraded despite receiving correct information from the user. The third case presents a mixed result. When the user identified the food as "Egg whites," the model correctly updated its fat estimation, resulting in an improvement. However, the estimations for Calories and Protein both declined. This suggests that while the model correctly associated "egg whites" with near-zero fat, its internal profile or subsequent quantity re-estimation for "egg whites" was less accurate than its baseline assumption (perhaps "whole eggs"). The new information, therefore, improved one metric while introducing significant error in others.
The most severe friction is observed in the fourth case. The user's answer, "Three" (strips of bacon), prompted a uniform decline across all metrics, making the final estimate significantly worse than the "vanilla" visual-only guess. Here, the model's internal database entry for "three strips of bacon" may be highly erroneous, and applying this flawed data point "poisoned" the entire meal calculation.

% The negative effect on nutrition value generation was expected as the resolution and timing of follow-up-question answers significantly impact the quality of generation, especially when the user themselves does not know the exact answer the model is searching for~\cite{srinivas2025substance}. The observed negative impact of the follow-up questions is likely compounded by the inherent noise in the human response data. The specific nature of the follow-up questions were non-trivial to annotate, requiring detailed subjective estimations not readily available from the image and on foods the study team did not cook or eat. Consequently, the annotators frequently struggled to answer these questions with real confidence, introducing a significant source of error or conflicting data into the prompt. Given the ambiguity of the required input, we believe this particular follow-up question evaluation on this constrained dataset is not truly representative of a real-world deployment where a user might be present at the time of eating.

A major limitation of this study was the necessary reliance on a small sample size for the follow-up question models due to the requirement for human input. Future research must prioritize gathering a significantly larger dataset of human-labeled follow-up-question responses to definitively determine the true potential of interactive estimation models and achieve tighter confidence intervals for all models. Ultimately, while the follow-up questioning technique showed promise for specific, clear-cut cases (Table~\ref{table:food_analysis_alt} rows 1 and 2), the most reliable immediate path to improved nutrition estimation lies in the synergistic combination of visual RAG context and explicit external data, as demonstrated by the robust performance of the RAG and receipt approach.
% separate the design of the tool from the evaluation of the tool
% 1. formative study 2. build a system (informed by formative study) 3. longitudinal user evaluation of system 4. technical evaluation 5. discussion of (3) and (4) 6. conclusion

% user study is shorter duration, because this is a medical thing, user satisfaction is not the only important thing. accuracy is also important. deep data using user data and synthetic data to benchmark key aspects and measurable perforamnce characteristics to measure user experience

% save discussion for after parts 3 and 4
% 

\section{Longitudinal User Evaluation}
\label{section:longitudinal}

While the preceding technical evaluation established SnappyMeal's strong foundation, validating its computational feasibility, these results are based on isolated performance metrics. 
Crucially, they do not account for the human factors critical to sustained dietary tracking, such as motivation, adherence, and the potential for technological fatigue over time. 
Therefore, to holistically assess how the system’s primary design principles of input flexibility and deep context-awareness translate into real-world usability and impact, we conducted a 3-week longitudinal study. 
This study transitioned our focus from the technical capabilities of the back-end architecture to the long-term changes in user behavior, adherence rates, and the evolving relationship between the user and the system's personalized conversational interface.

\subsection{Methods}

\subsection{Participants and Recruitment}
Recruitment was conducted via word-of-mouth within a university setting and a formal advertisement announcement posted in university departments' official Slack channel and mailing lists. To encourage completion and mitigate attrition, participants were offered a compensation of one \$20 USD electronic gift card for each full week of study completion, totaling \$60 for the entire 3-week period.
A total of 12 eligible participants were initially recruited for this longitudinal study.
Of the eligible participants, 8 individuals downloaded the app, and the final sample consisted of 6 participants who completed the full 3-week study period. The participant pool primarily comprised individuals in the 18-24 age range, consistent with the recruitment strategy. 4 of the 6 participants reported prior experience logging food.

\begin{table}[h]
    \centering
    \caption{Characteristics of 3-Week Study Participants}
    \label{tab:participants_longitudinal}
    \begin{tabular}{ccccc}
        \toprule
        \textbf{Participant ID} & \textbf{Age Range (Years)} & \textbf{Sex} & \textbf{Tracking History} & \textbf{Occupation} \\
        \midrule
        P1 & 18-24 & M & 6 months to 1 year & Student \\
        P2 & 18-24 & F & 1-6 months & Engineer \\
        P3 & 18-24 & M & Less than 1 month & Software Engineer \\
        P4 & 18-24 & F & 1-6 months & Student  \\
        P5 & 25-34 & F & 1-6 months & Student \\
        P6 & 18-24 & M & 6 months to 1 year & Student \\
        \bottomrule
    \end{tabular}
\end{table}

\subsubsection{Inclusion/Exclusion Criteria} Inclusion criteria for participation included smartphone ownership and a commitment to consistently track dietary intake throughout the study. Exclusion criteria were applied to individuals with a self-reported history of disordered eating or diagnosed eating disorders, as the nature of the study could pose a potential health risk.

\subsubsection{Ethics}
A detailed study protocol was submitted to and approved by the IRB at the host institution for this study prior to recruiting participants. 

\subsection{Experiment Setup}

The study was designed as a 3-week longitudinal study to evaluate the performance and user experience of the novel nutrition tracking app. The experiment was conducted in a real-world, naturalistic setting, with participants using the app in their daily lives.

To prevent system overload and manage data collection efficiently, the participation was staggered. Participants were onboarded in small groups at different times, ensuring that the backend infrastructure—including the Flask server and PostgreSQL database—could handle the concurrent requests without crashing. This allowed for a smooth data collection process, particularly for the multimodal inputs (images and conversational data).

\subsection{Study Procedure}

The study was not divided into distinct phases but rather operated as a continuous, 3-week active tracking period for each participant.

\subsubsection{Onboarding} Participants created an account, with basic demographic information and initial nutrition goals.

\subsubsection{Active Tracking} For the 3-week duration, participants were instructed to use the app to log all meals, snacks, and beverages. They were encouraged to utilize the app's multimodal features, including text, image, and audio inputs, as the primary method for logging.

\subsubsection{Data Collection} The system automatically extracted and saved all raw user data. This included:

\begin{itemize}
    \item \textbf{Conversational Data:} All interactions with the Gemini API were stored in the conversations table.

\item \textbf{Food Tracking Data:} The raw media files (images, audio) and their corresponding estimated nutritional logs were stored in the food\_logs table, with media files offloaded to the MinIO file bucket.
\end{itemize}

\subsection{Challenges and Data Consistency}

A key challenge observed during the study was the low consistency in participants' food tracking habits. This resulted in a dataset with significant variability in the number of logged meals per day and the thoroughness of the logs. This finding will be addressed in the discussion section, as it highlights a common hurdle in longitudinal nutrition tracking studies and provides a realistic context for the app's performance. The study focused on extracting and analyzing the available raw data to understand user behavior and system performance under real-world usage conditions, even with inconsistent input.

\subsection{Results}

\subsubsection{Survey results}
An exit interview form was sent to the 6 participants.
From these interviews, we found high engagement for most users, suggesting the process was manageable for many, despite some frustration.

% \textcolor{red}{TODO: move exit interview survey results here}

A section of the form was dedicated to answering participants' agreement to some statements with answers 1 to 5 with 1 meaning "Strongly Disagree" and 5 meaning "Strongly Agree." Participants reported that the app made them feel more aware of their eating habits ($\bar{x} = 3.67$). Additionally, participants found the follow up questions related relevant to the food they were uploading ($\bar{x} = 3.83$) and relevant to the goals they were seeking to achieve ($\bar{x} = 3.83$). When it came to the systems' usability, participants appreciated the ability to edit their logs ($\bar{x} = 4.33$).

\subsubsection{User engagement}
Engagement with the application was generally high during the three-week trial, with four of six participants reporting daily logging (7 days per week), one logging 5–6 days per week, and one logging 1–2 days per week. However, the estimated time required for an individual food log showed high variance, ranging from under one minute ($n=2$) to over five minutes ($n=1$). Outside of technical failures, the primary self-reported reasons for missed logging were that the process felt too time-consuming or cumbersome ($n=3$) or forgetting to log ($n=3$).

\begin{figure*}[ht!]
\centering
\begin{subfigure}{.4\textwidth}
    \centering
    \includegraphics[width=\linewidth]{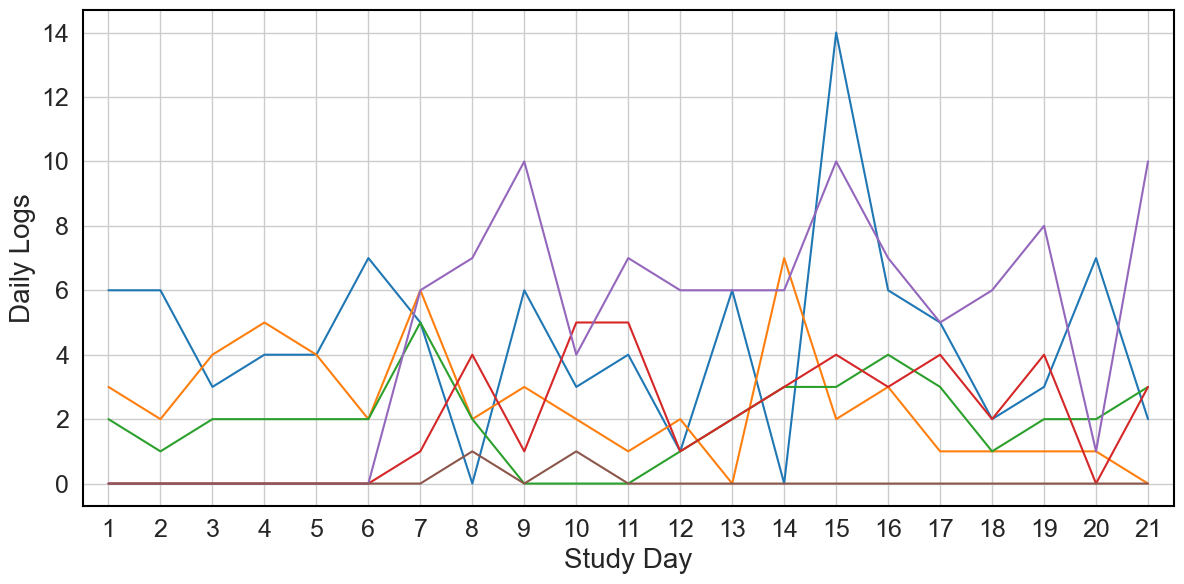}
    \caption{Frequency of user logs over time.}
    \label{fig:logs_over_time}
\end{subfigure}
\begin{subfigure}{.5\textwidth}
    \centering
    \includegraphics[width=\linewidth]{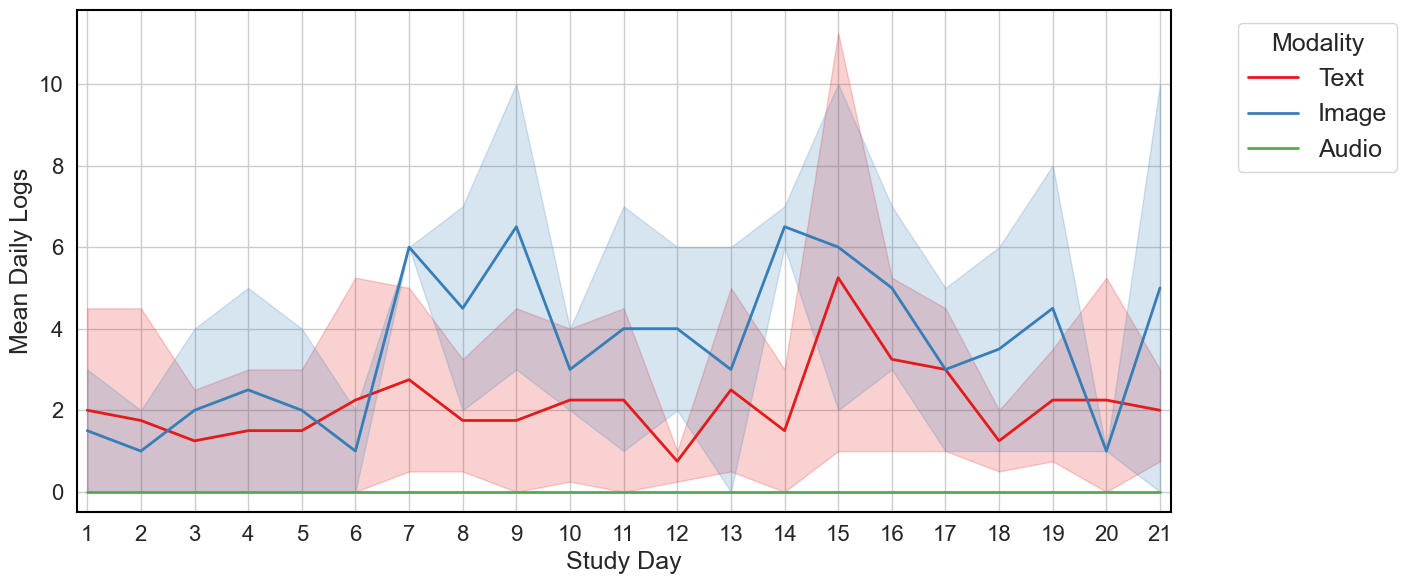}
    \caption{Mean frequency of modalities logged over time.}
    \label{fig:modalities_over_time}
\end{subfigure}

\caption{User logs over time.}
\label{fig:logs_over_time_joint} 
\Description[The frequency of user logs over time for each of the 8 users as well as the modalities over time.]{The first image depicts frequency of user logs over times for each of the 8 users. Even through the inconsistent frequency, we can see that there is no huge drop off over time. The second image depicts mean frequency of user logs, separated by modality. We see that the audio is always at a flat line of 0. The image and text modalities have a little bit less of a consistent trend where they start low and increase over time.}
\end{figure*}

All users completed the 21 days of data collection. Figure~\ref{fig:logs_over_time} a shows the daily number of log entries per user over the course of the study. With the exception of one user who only entered two logs, users generally logged their food multiple times per day. Participants generally had a high variability in logging frequency, with some users forgetting to log on some days and compensating by logging extra on other days. 

Fig~\ref{fig:modalities_over_time} shows the number of logs by modality over the course of the study. There was a robust mix of modalities, with many users preferring images. Interestingly, we note that no users chose to use audio. This could be because public settings such as restaurants are not conducive to recording audio, or they were eating while logging. We note that our participants represent a younger demographic. Older users less familiar with technology may be more inclined to use audio inputs.

Preference for logging method was split, with three participants preferring Image Logging and three preferring Text Logging for efficiency. However, the image recognition component was a source of friction. One participant rated the accuracy as "Sometimes accurate (25-50\% of the time)." The results highlight a significant disconnect between the perceived benefit of the application—increased awareness and data visualization—and the core interaction cost associated with the logging process.

\begin{figure*}[t!]
    \centering
    \includegraphics[width=1\linewidth]{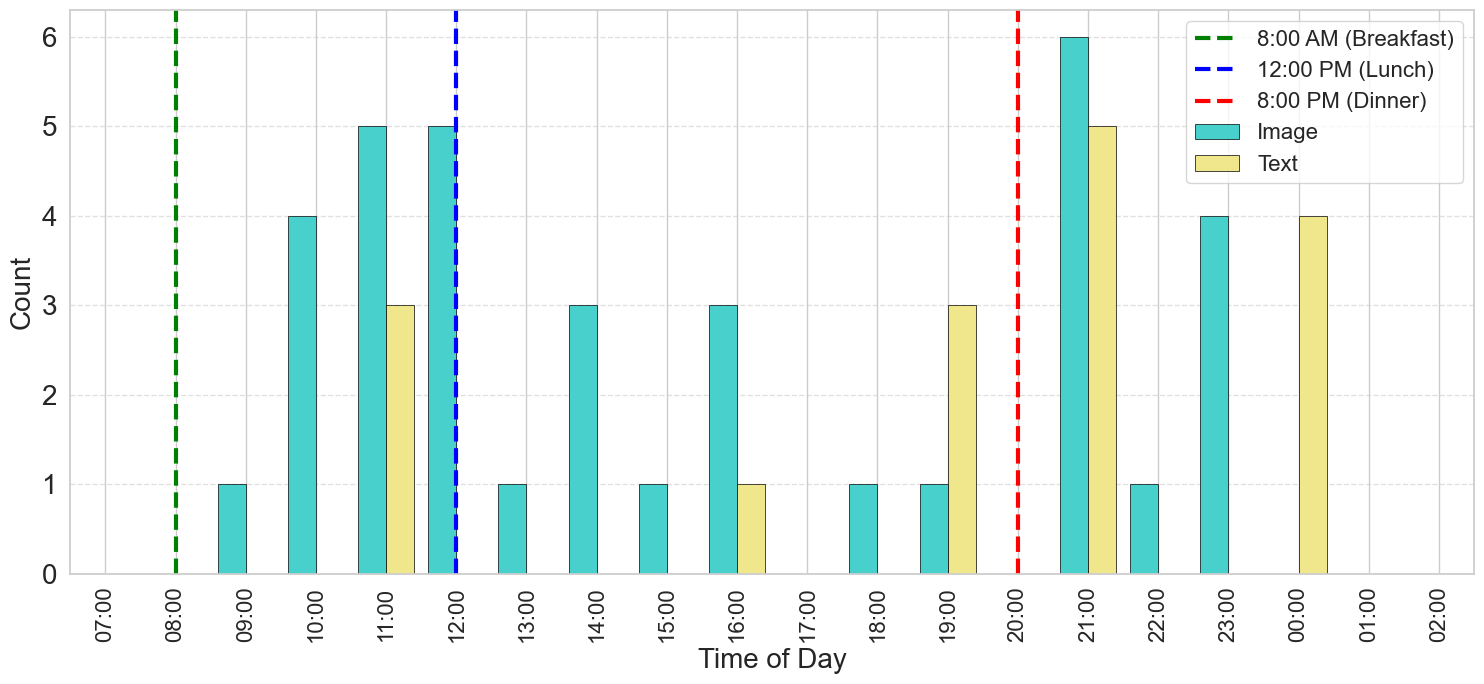}
    \caption{One day timeline showing the mix of modalities used throughout the day}
    \label{fig:timeline}
    \Description[This image depicts one user's logs and their modalities used against time of day. Images are less frequent outside of regular meal times.]{This image depicts one user's logs and their modalities used against time of day. The image also has lines depicting the times of standard meals for breakfast, lunch, and dinner. It appears as if images are more frequent at times further from regular meal hours.}
\end{figure*}

In addition to these high level trends, we also observe that users generally prefer a mix of modalities. Fig~\ref{fig:timeline} shows a timeline representative of an average day for one user. This user seems to prefer logging with image around actual meal times while preferring to log with text for snack times.
These figures show that user preferences for how the enter data varies and our application accommodates this to reduce the mental load of tracking.

\subsubsection{Interactive AI follow up}
\begin{table}[ht!]
\caption{Gemini Classification of follow up questions classified. Definitions for categories and the prompt used to classify can be found in Appendix \ref{appendix:classification_prompt}.}
\label{table:follow_up_categories}
\begin{tabular}{lrr}
\toprule
 Category & Percentage ($\%)$ & Count ($n$) \\
\midrule
Quantity \& Portion Size & 34.3 & 610\\
Food Type \& Detail & 31.8 & 566\\
Preparation \& Source & 20.1 & 357\\
Consumption Ratio & 12.4 & 221\\
Other & 1.4 & 25\\
\bottomrule
\end{tabular}
\end{table}

\begin{figure*}[t!]
    \centering
    \includegraphics[width=1\linewidth]{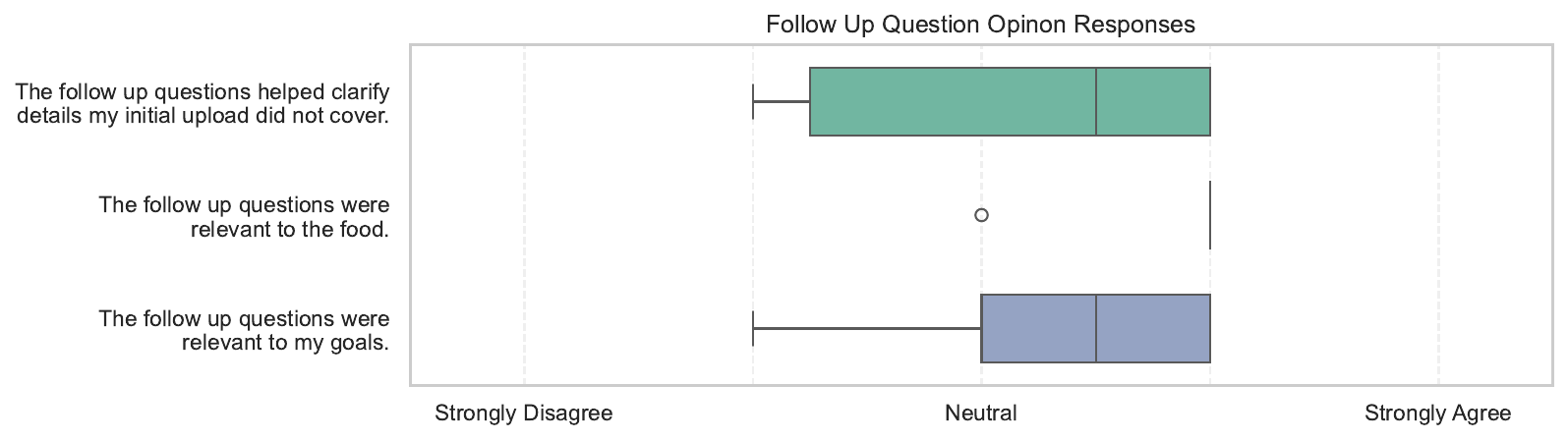}
    \caption{Responses to "follow up"-focused questions in exit survey.}
    \label{fig:follow_up_q_responses}
    \Description[A likert scale bar chart for three questions about follow up questions.]{A likert scale bar chart for three questions about follow up questions. Users tended to agree that the follow up questions helped clarify details their initial uploads did not cover, the follow up questions were relevant to the food, and the follow up questions were relevant to their goals.}
\end{figure*}

Follow-up questions were a generally new concept in food logging for the participants. Fig \ref{fig:follow_up_q_responses} reveals participants generally believed the follow-up questions were relevant to their food and personal goals as well as helped clarify details their initial uploads did not cover. Table~\ref{table:follow_up_categories} exhibits the distribution of genre of follow up questions with most of the questions being about quantity \& portion size and food type \& detail, questions that were important to clarify in Section~\ref{section:formative_findings}.

\begin{figure*}[t!]
    \centering
    \includegraphics[width=1\linewidth]{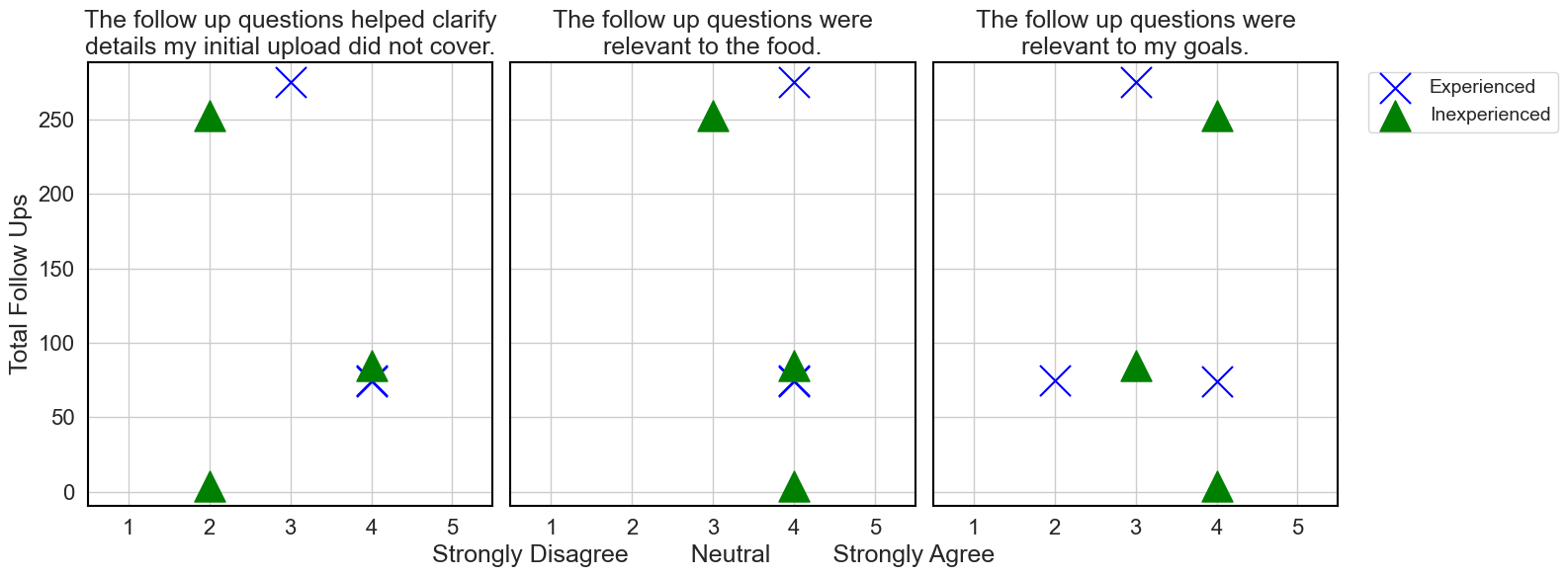}
    \caption{}
\label{fig:likert_vs_follow_up_count}
\Description[Three scatter plots demonstrating a user's total number of received follow up questions against their agreement with statements about follow up questions.]{Three scatter plots demonstrating a user's total number of received follow up questions against their agreement with statements about follow up questions. The statements were "the follow up questions helped clarify details my initial uploads did not cover," "the follow up questions were relevant to the food," and "the follow up questions were relevant to my goals."}
\end{figure*}

\begin{figure}
    \begin{subfigure}{.45\textwidth}
        \centering
        \includegraphics[width=\linewidth]{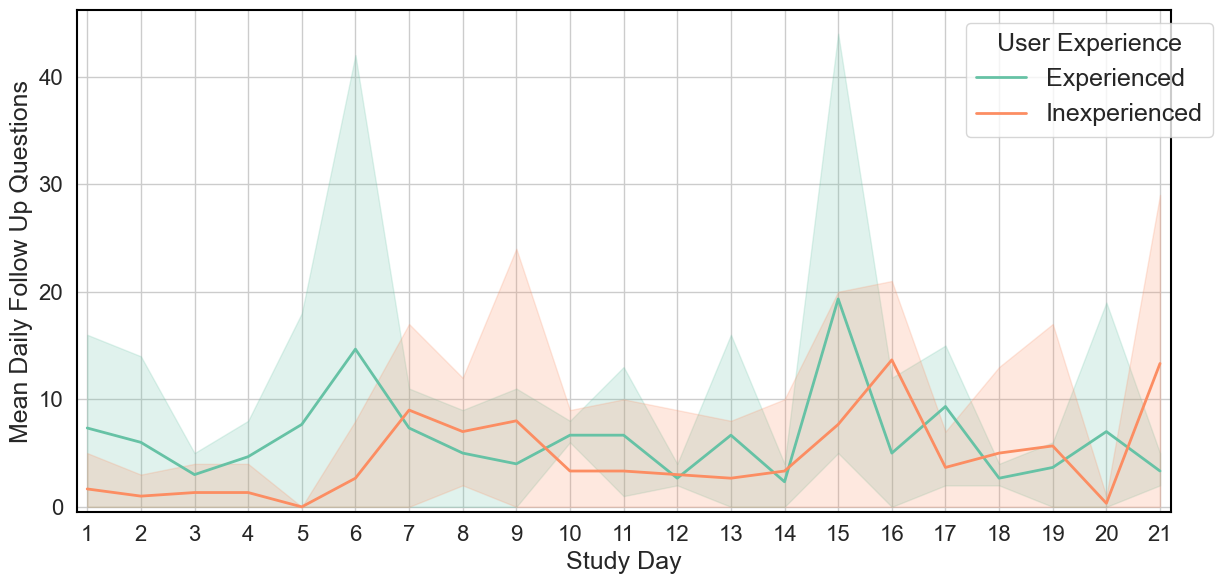}
        \caption{Mean daily follow up questions over time.}
    \label{fig:follow_up_questions_over_time_experienced}
    \Description[The mean number of daily follow up questions asked to users separated by whether the user is experienced.]{The mean number of daily follow up questions asked to users with two lines, one for experienced users and one for inexperienced users.}
    \end{subfigure}
    \hfill 
    \begin{subfigure}{.45\textwidth}
        \centering
        \includegraphics[width=\linewidth]{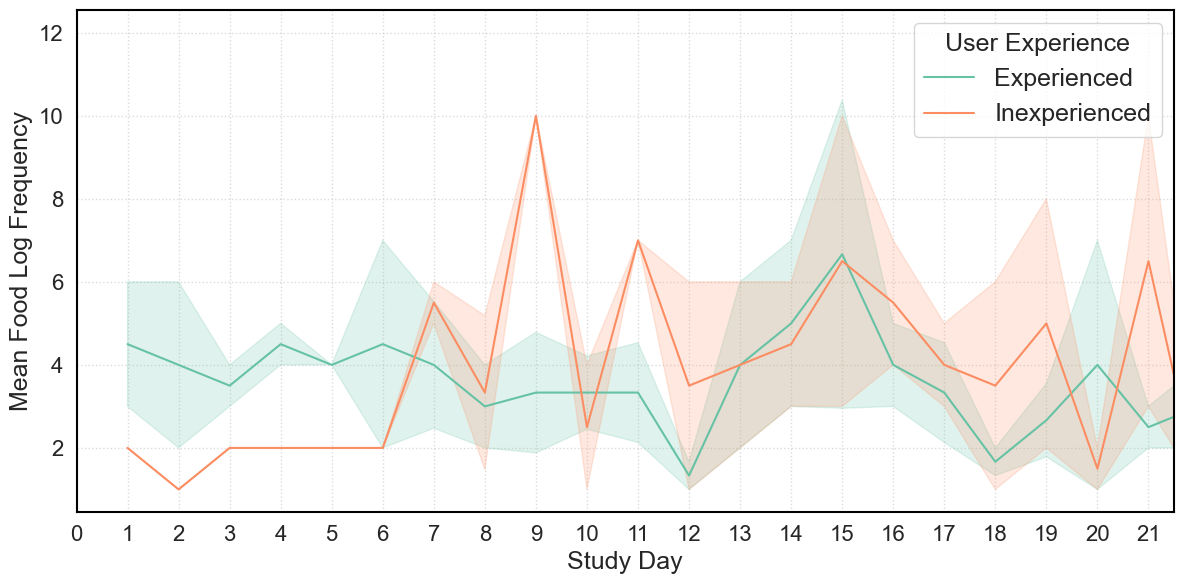}
        \caption{Mean daily user food logs over time.}
        \label{fig:experienced_mean_logs}
        \Description[The mean number of daily food logs by users separated by whether the user is experienced.]{The mean number of daily food logs by users with two lines, one for experienced users and one for inexperienced users.}
    \end{subfigure}
    \label{fig:experienced_logs}
    \caption{Food logging trends by user experience (more than 6 months of experience with nutrition tracking).}
\end{figure}

While the follow-up questions were rated as relevant to the food, the quantitative results show they did not make the process easier than traditional methods. 
The feedback suggested the follow-up questions failed to adapt to input context. 
For instance, a user employing Text Logging reported receiving follow-up questions "phrased for photo input," and another complained of generic, frustrating defaults: "...it would default to 'how much chicken did you bake' which was frustrating to edit." 
Furthermore, users requested the ability to skip follow-up questions for simple, single-ingredient foods (e.g., a banana), emphasizing the need for greater efficiency.
Fig~\ref{fig:likert_vs_follow_up_count} reveals there was no clear correlation between the number of follow-up questions received and user's opinion reflecting the follow-up question methodology. This suggests that the survey results were not significantly biased by the number of follow-up questions administered.
However, Fig~\ref{fig:follow_up_questions_over_time_experienced} reveals more experienced food loggers (at least 6 months of experience) generally received more follow-up questions. 
This is likely because more experienced loggers tended to log more in general, as illustrated in Fig~\ref{fig:experienced_mean_logs}.

The follow-up questions were intended to improve data quality and self-awareness when logging food, but they inadvertently became a source of user friction. While they were rated as relevant to the food, the quantitative results show follow-up questions did not make the process easier than traditional methods. The feedback suggested the follow-up questions failed to adapt to input context. For instance, a user employing Text Logging reported receiving follow-up questions "phrased for photo input," and another complained of generic, frustrating defaults: "...it would default to 'how much chicken did you bake' which was frustrating to edit." Furthermore, users requested the ability to skip follow-up questions for simple, single-ingredient foods (e.g., a banana), emphasizing the need for greater efficiency.

 Beyond the exit interviews, analysis of the application's usage data provided quantitative metrics on logging accuracy and behavior. Due to the app's configuration, we tracked the frequency of log edits and deletions. Out of 502 total food logs, 104 (20.7\%) were edited and 29 (5.8\%) were deleted.

\subsection{Longitudinal Study Discussion}
\label{section:longitudinal_discussion}

The 21-day longitudinal deployment successfully evaluated the novel nutrition tracking application in a real-world setting, providing critical insights into the trade-off between minimizing tracking fatigue and maintaining data accuracy.
The study demonstrates two significant positive outcomes: (1) The system successfully fostered high user engagement and adherence over the 21-day trial, and (2) it achieved its primary goal of increasing users' self-awareness of their eating habits.

%\subsubsection{}

Engagement was notably high, with all six participants completing the entire 21-day study and four of six logging daily. This high adherence, despite self-reported frustrations with the time required ($n=3$), suggests that the perceived benefits of the system—namely, feeling more aware of their eating habits ($\bar{x} = 3.67$)—were compelling enough to overcome usability hurdles. This finding is a key success, as "logging fatigue" and user drop-off are primary challenges for longitudinal health apps~\cite{kyoung2014understanding}.

Despite these successes, our findings also reveal a critical tension: the very AI features intended to enhance the logging experience also became new sources of friction. The discussion section below will unpack this by examining (1) the success and challenges of the AI-driven follow-ups, (2) the value and friction of the multimodal design, and (3) the implications for designing future context-aware food logging systems.

% \subsubsection{AI Follow-ups}

A key success of our system was the content and relevance of the AI-driven follow-ups. Participants strongly agreed that the questions were relevant to their food ($\bar{x} = 3.83$) and their personal goals ($\bar{x} = 3.83$). The system demonstrated its utility by helping users "clarify details their initial uploads did not cover" (Fig~\ref{fig:follow_up_q_responses}). Our quantitative analysis confirms this, showing the AI correctly targeted the most critical missing information: "Quantity \& Portion Size" (34.3\%) and "Food Type \& Detail" (31.8\%) (Table~\ref{table:follow_up_categories}). This demonstrates the potential of an interactive AI to intelligently guide users toward higher-quality data collection.

\subsubsection{Limitations}
Several limitations must be considered when interpreting these findings. First, our sample size ($N=6$) is small and relatively homogeneous, limiting the generalizability of our results. The speculation that "older demographics" might use audio remains untested and the majority of participants were tech-savvy students or engineers. This demographic is generally characterized by a high degree of technological aptitude and comfort with new applications and systems. The reported high engagement and perceived manageability of the logging process may be partially attributable to this. 

Second, the 21-day study duration was sufficient to observe sustained use without significant drop-off (Fig~\ref{fig:logs_over_time_joint}), but a longer longitudinal study would be needed to assess the long-term effects of "logging fatigue." Finally, our metrics for "logging burden" were based on self-report; future studies could benefit from more objective measures, such as timestamped interaction logs to precisely quantify the time spent on each log and edit.

In summary, the longitudinal study yielded positive indicators of high user engagement and perceived awareness benefits among a tech-savvy cohort. 
The critical and specific feedback regarding contextual failure in the follow-up questions and friction in image recognition offers a powerful set of instructions for future iterations, allowing us to focus on dramatically lowering the interaction cost to further maximize compliance and data quality.
Ultimately, the goal is to make the system proactively adaptive rather than merely reactive.

\section{Discussion}

Our formative formative study revealed that dietitians and food journalers prioritize flexible, contextually-aware data capture. 
While many of the model features occurred in the backend, follow-up questions allowed users to focus on the experience of eating.
Though this feature resulted in a decreased accuracy during our system evaluation (Section~\ref{section:system_eval}), the longitudinal study (Section~\ref{section:longitudinal}) demonstrated its significant value to real-world user, who appreciated the added awareness it provided.

\subsection{AI Follow ups}

While challenging to quantify, the follow-up questions improved user engagement. Users reported feeling more involved in the food logging process. 
The act of being asked served as a behavioral intervention, driving the perceived value in the longitudinal study.
Despite this, the follow ups resulted in user effort that led to statistically noisier data, sometimes due to the poor nutrition annotations of users. 
The negative effect of textual follow-ups on nutrition value generation was expected as the resolution and timing of follow-up-question answers significantly impact the quality of generation, especially when the user themselves does not know the exact answer the model is searching for~\cite{srinivas2025substance}.
The observed negative impact of the follow-up questions is likely compounded by the inherent noise in the human response data. 
The specific nature of the follow-up questions were non-trivial to annotate, requiring detailed subjective estimations not readily available from the image and on foods the study team did not cook or eat. 
Consequently, the system evaluation annotators frequently struggled to answer these questions with real confidence, introducing a significant source of error or conflicting data into the prompt.
Given the ambiguity of the required input, we believe this particular follow-up question evaluation on this constrained dataset is not truly representative of a real-world deployment where a user might be present at the time of eating.

However, this success in \textit{what} to ask was disconnected from \textit{how} and \textit{when} to ask.
This "context-insensitivity" was a primary source of user frustration. 
The main complaints included:
lack of modality awareness, generic and inefficient defaults, and lack of adaptive friction.
This finding has significant implications for HCI: an "intelligent" system is not just one that identifies correct information, but one that knows when not to intervene. 
The lack of correlation between the number of follow-ups and user opinion (Fig~\ref{fig:likert_vs_follow_up_count}) reinforces this; it was not the quantity of questions that mattered, but the quality and context of each interruption. 

Future systems should be trained to actively default to user convenience and only introduce friction when the benefit of the improved estimation significantly outweighs the burden on the user.
This necessitates a new set of system intervention thresholds based on user-centric criteria, ultimately leading to more usable and trustworthy automated logging tools.

\subsection{User-Driven Flexibility}
The multimodal input design was a clear success in providing user-driven flexibility. 
Participants organically developed logging strategies to fit their daily context. 
As visualized in Fig~\ref{fig:timeline}, a user might prefer images for complex meals (e.g., at mealtimes) and text for simple snacks. 
This accommodation for varying user preferences and situations is a positive outcome, demonstrating that the system successfully reduced the mental load of tracking by allowing users to choose the most efficient path. 
This benefit was, however, hampered by the failure of image recognition. While there was a preference for image logging ($n=3$), it was directly undermined by the poor performance of the image recognition model ("Sometimes accurate (25-50\%)"). 
The high "edit" rate (20.7\% of all logs) and the corresponding high appreciation for the edit feature ($\bar{x} = 4.33$) are symptoms of this system making frequent mistakes.

In the future, systems should support more modalities such as inferences from wearable metrics and more personalized meta-data. 
Time of day, location, and other meta-data can assist the inference of models. For example, if the user's location is at a restaurant a system could automatically retrieve an online menu to supplement its context. With the wide access to user data~\cite{istepanian2018mhealth}, systems can now perform extrapolations that would improve the accuracy of nutrition estimates and infer diet choice consequences for each user.

\subsection{Moving from "Interactive" to "Context-Aware"} 
A core lesson from this study is the need to shift from an interactive system to a context-aware one. The current system places the burden of context-switching and error correction on the user. Based on our findings, we propose two key design implications for future work:
\begin{enumerate}
\item \textbf{Pipe Modality Context to the AI:} The AI model (Gemini) must be explicitly prompted with the input modality. This would prevent basic errors like asking photo-phrased questions for text logs and represents a straightforward, high-impact fix.
\item \textbf{Implement Adaptive Friction:} The system should dynamically adjust its level of intervention based on the confidence of the initial log. For a high-confidence text log ("banana") or a high-accuracy image recognition, the system should default to no follow-up question to maximize efficiency.
\end{enumerate}
System intelligence must practice a restraint. Future work should focus on building context engines that manage the \textit{when} and \textit{how} of user intervention, allowing users to concentrate on their primary tasks rather than correcting an overzealous digital assistant. This paradigm shift will lead to logging tools that are not only accurate but truly supportive of real-world behavior change.

\subsection{Implications For Future Work}

In summary, the design and evaluation of SnappyMeal revealed a fundamental tension in the development of automated logging tools: the conflict between algorithmic accuracy and user-perceived value. Our system evaluation demonstrated that follow-up questions, while intending to gather crucial missing data, introduced noise and cognitive load that reduced the final accuracy of the nutrition estimates. However, the subsequent longitudinal user study demonstrated that this very friction was valuable, acting as a prompt for user self-contextualization and reflection\textemdash{}the true goal of many behavior change applications.

This outcome necessitates a paradigm shift from building merely "interactive" systems to engineering truly "context-aware" systems that practice restraint. Our key design implication is the implementation of a system's decision to intervene being governed by the input modality, the confidence of the initial log, and the real-world cost of user interruption. This system intelligence must be rooted not only in what information is missing but in when and how to ask for it.

The challenges we identified extend beyond nutrition directly into other high-stakes, data-intensive domains. Any task where an LLM needs to make a high-stakes decision based on incomplete user input will require it to actively probe for necessary details rather than relying solely on the initially logged information. This extends to clinical settings as well. \citet{englhardt2024from} find that adding more data to their prompt, even if the data is accurate, does not necessarily improve LLM reasoning performance on extrapolating depression and anxiety from activity, sleep and social interaction data. The study emphasizes that the LLMs can generate rigorous analysis and natural language insights for clinicians, but the "improvement" does not necessarily come from logging more data, but rather creating a system that can contextualize and interpret the logged data in a clinically useful manner. More specifically, \citet{li2024beyond} note the importance of follow-up questions clinical pre-consultations and its necessity in systems built to automate these processes. The improvement in these systems comes from designing frameworks, such as follow-up questions, that allow models to ask "good" questions based on fine-grained attributes. This ability to generate relevant, targeted follow-up questions effectively improves the quality of the logged/collected data during the interaction, reducing the need for extensive provider communication later. All in all, the key to improving LLM performance lies in designing proactive, context-aware systems, specifically by integrating frameworks for generating targeted follow-up questions, that effectively refine and contextualize the input, thereby enhancing data quality and the clinical utility of the analysis.

\section{Conclusion}

While dietary tracking is critical for understanding health outcomes, current methods like app-based journaling are inflexible, resulting in poor user adherence and imprecise nutritional summaries. 
SnappyMeal, our proposed AI system, enhances data quality by intelligently posing goal-dependent follow-up questions to acquire missing context and by utilizing information retrieval from grocery receipts and nutritional databases.
We validated SnappyMeal through public benchmarks and a 3-week, in-the-wild deployment.
Participants reported high satisfaction with the multiple input methods and strong perceived accuracy, a sentiment supported by objective benchmark performance.
These findings suggest that multimodal AI systems can substantially improve adherence and accuracy, heralding a new class of intelligent self-tracking applications.
Ultimately, SnappyMeal demonstrates the need for restrained intelligence.
The most usable and trustworthy logging tools will be those that prioritize user convenience by defaulting to maximum efficiency and only intervening when the expected benefit to the estimate significantly outweighs the cognitive cost to the human.
\bibliographystyle{ACM-Reference-Format}
%\bibliography{sample-base}
\bibliography{references}
\newpage
\appendix
\section{Database Schema}
\label{appendix:food_logging_relation_schema}
\begin{figure*}[ht!]
    \centering
    \includegraphics[scale=0.6]{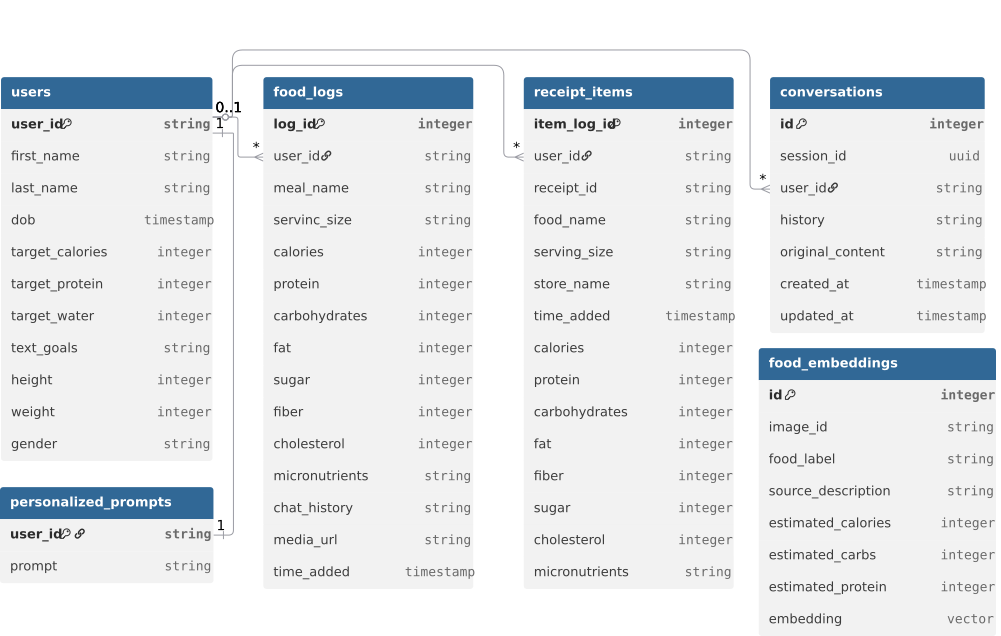}
    \caption{Relational Database Schema}
    \Description[The relational database schema providing visualization of the users, food\_logs, personalized\_prompts, receipt\_items, conversations, and food\_embeddings tables.]{The relational database schema providing visualization of the users, food\_logs, personalized\_prompts, receipt\_items, conversations, and food\_embeddings tables.}
\end{figure*}

\newpage 
\section{Prompts}

\subsection{Vanilla Food Log Generation Prompt}
\label{appendix:vanilla_prompt}
\texttt{Analyze the nutritional content of the food items in this piece of media. The media can be image, text, or audio.\\
You are also provided with a chat history that clarifies potentially missing information.\\
Please identify the main ingredients and estimate the approximate macronutrient breakdown (calories, protein, carbohydrates, fat) and highlight any significant micronutrients that are visually discernible.\\
Output the results in a structured JSON format with the following keys: "ingredients", "macronutrients", and "micronutrients". \\
Under "meal\_name", come up with a name that summarizes the food in as few words as possible.\\
Under "ingredients", list all the ingredients that could affect the nutritional value. \\
Under "serving\_size", estimate how much food there is in customary units (oz, cups, tbsp, etc.).\\
Under "calories", estimate how many calories are in the food. Don't provide the units, use the most standard unit.\\
Under "protein", estimate how much protein is in the food. Don't provide the units, use the most standard unit.\\
Under "carbohydrates", estimate how many carbohydrates are in the food. Don't provide the units, use the most standard unit.\\
Under "fat", estimate how much fat is in the food. Don't provide the units, use the most standard unit.\\
Under "fiber", estimate how much fiber is in the food. Don't provide the units, use the most standard unit.\\
Under "sugar", estimate how much sugar is in the food. Don't provide the units, use the most standard unit.\\
Under "cholesterol", estimate how much cholesterol is in the food. Don't provide the units, use the most standard unit.\\
Under "micronutrients", list any significant vitamins or minerals that are visible in the image and their amounts. Only include one number and nothing else. If you think it's between some range, use the average. Don't include units.\\
Reminder that JSON formatting requires double quotes.\\
Example output:\\
\{example\_output\}\\
Do not output anything except the JSON. 
}

\colorlet{punct}{red!60!black}
\definecolor{delim}{RGB}{20,105,176}
\colorlet{numb}{magenta!60!black}
\lstdefinelanguage{json}{
    basicstyle=\normalfont\ttfamily,
    numbers=left,
    numberstyle=\scriptsize,
    stepnumber=1,
    numbersep=8pt,
    showstringspaces=false,
    breaklines=true,
    frame=lines,
    literate=
     *{0}{{{\color{numb}0}}}{1}
      {1}{{{\color{numb}1}}}{1}
      {2}{{{\color{numb}2}}}{1}
      {3}{{{\color{numb}3}}}{1}
      {4}{{{\color{numb}4}}}{1}
      {5}{{{\color{numb}5}}}{1}
      {6}{{{\color{numb}6}}}{1}
      {7}{{{\color{numb}7}}}{1}
      {8}{{{\color{numb}8}}}{1}
      {9}{{{\color{numb}9}}}{1}
      {:}{{{\color{punct}{:}}}}{1}
      {,}{{{\color{punct}{,}}}}{1}
      {\{}{{{\color{delim}{\{}}}}{1}
      {\}}{{{\color{delim}{\}}}}}{1}
      {[}{{{\color{delim}{[}}}}{1}
      {]}{{{\color{delim}{]}}}}{1},
}

\newpage
\subsection{Example Generated Food Log JSON}
\begin{lstlisting}[language=json,firstnumber=1]
{
    "meal_name": "Peanut butter and celery",
    "ingredients": ["peanut butter", "celery"], 
    "serving_size": "1 large celery stalk with 2 tablespoons creamy peanut butter",
    "meal_type": "snack",
    "date": "2025-05-07T10:13:27Z",
    "calories": 280,
    "protein": 11,
    "carbohydrates": 16,
    "fat": 20,
    "fiber": 4,
    "sugar": 7,
    "saturated_fat": 4,
    "cholesterol": 0,
    "micronutrients": 
    {
        "vitamin_k_mcg": 30,
        "vitamin_a_iu": 500,
        "folate_mcg": 40,
        "potassium_mg": 450,
        "magnesium_mg": 60,
        "phosphorus_mg": 120,
        "vitamin_e_mg": 2,
        "niacin_mg": 3,
        "zinc_mg": 1
    }
}  
\end{lstlisting}

\newpage
\subsection{Follow Up Question Generation Prompt}
\label{appendix:follow_up_question_generation_prompt}
\texttt{You are an AI assistant for a food logging service. Your main goal is to gather detailed information about a user's meal to create a complete food log entry. Users will provide media (image, text, or audio) of their food.\\
\#\#\# Task Workflow\\
1.  **Analyze Input:** Examine the user's media to identify the food and any information already provided.\\
2.  **Check for Missing Information:** Compare the extracted information against all required fields for a food log entry: `serving\_size`, `calories`, `protein`, `carbohydrates`, `fat`, `sugar`, `fiber`, `cholesterol`, and `micronutrients`.\\
3.  **Formulate a Single Clarifying Question:** If any crucial information is missing or ambiguous, you must formulate **one** concise, direct follow-up question.\\
    * Prioritize questions that fill the most critical gaps first (e.g., what the food is, then the quantity, then preparation method).\\
    * The question must be relevant to the user's input. For example, do not ask about an image if one was not provided.\\
    * For homemade meals, if possible, refer to recent receipts to ask a targeted question about ingredients.\\
    * Avoid greetings, conversational filler, or explanations. The output should be **only** the question.\\
    * The question should directly help gather information for the food log fields. Do not ask about nutritional measurements (e.g., "How many calories?")\\
    * **Portion Size:** Try to understand how much food the person ate (e.g. "How many slices of toast were there?"). \\
4.  **Determine Question Type:** The question must be one of the following types:\\
    * **text:** Answered with free-form text.\\
    * **select:** Answered by selecting one option from a list. This is the preferred type and should have no more than 3 options.\\
5.  **Output Format:** If a question is needed, output only the question, its type, and any options (if applicable). These three pieces of information must be separated by a semicolon (;).\\
    * Example: `How many slices of pizza did you eat?;select;[1,2,3,4,5]`\\
    * Example: `Are there any unseen ingredients in the lasagna?;select;[yes,no]`\\
    * Example: `What is inside your burrito?;text;[]`\\
\#\#\# Constraints\\
* **One Question Only:** Generate a single question at a time.\\
* **Relevance:** All questions must be directly relevant to completing the food log fields.\\
* **No Timestamps:** You are forbidden from asking about the time the meal was eaten.\\
* **No Extraneous Text:** Do not output any text other than the formatted question.\\
* **Clarity on Units:** For `select` questions, explicitly state the units (e.g., "How glasses of milk?"). Do not ask vague questions like "What is the serving size?" and do not ask about measurable units like "How many grams of protein?". Remember, it's hard for individuals to guess portion sizes.\\
* **Receipts:** Only ask questions about receipts if the meal seems to contain something from the receipt list and the question is relevant.\\
* **Necessity:** Do not ask unnecessary questions. If you can determine a value or fact, don't ask about it. If the value or fact is hard for a human to estimate, don't ask about it.\\
Example output:\\
What percentage of the food did you consume?;text;[]\\
Example output:\\
What is inside of your burrito?;text;[]\\
Example output:\\
Are there any unseen ingredients in the lasagna?;select;[yes,no]\\
Example output:\\
Is this curry homemade?;select;[yes,no]\\
Example output:\\
How was the chicken cooked?;select;[roasted,fried,other]\\
Example output:\\
How were the vegetables prepared?;select;[stir fried,steamed,raw,other]\\
Example output:\\
How many slices of pizza did you eat?;select;[1,2,3,4,5,6,7,8,9,10]\\
Example output:\\
Was this protein shake store bought or homemade?;select;[store bough,homemade]\\
Example output:\\
Where did you buy your protein powder from?;text;[]\\
Example output:\\
Is this the Chobani yogurt you bought from Safeway?;select;[yes,no]\\
}

\newpage
\subsection{Follow Up Question Classification Prompt}
\label{appendix:classification_prompt}
\texttt{You have been tasked with classifying questions surrounding nutrition tracking.\\
There are four categories:\\
1. Preparation \& Source: This category focuses on how the food was prepared or where it came from.\\
2. Food Type \& Detail: These questions seek specific detail about a particular ingredient or food item, useful for precise nutrient tracking.\\
3. Quantity \& Portion Size: This is for tracking the absolute amount or portion of a specific food item consumed.\\
4. Consumption Ratio: This category is for questions that gauge the user's portion of a larger, visible meal.\\
You must output a JSON containing the fields "question" and "category". Only output this JSON with no other formatting.\\
If you think the question does not belong to any of the aforementioned categories, you can set the "category" to "None".\\
Here is the question:}

\newpage
\section{Equations}
\label{appendix:equations}

\subsection{Error Formulae}
\label{appendix:equations-error}
$\hat{y}$ denotes an estimated value while $y$ denotes an observed or ground-truth value. $N$ represents the number of samples being evaluated.
\begin{align}
    \text{MAE} =& \frac{1}{N} \sum_{i=1}^{N} | \hat{y}_i - y_i | \label{equation:mae} \\
    % \text{MSE} =& \frac{1}{N} \sum_{i=1}^{N} (\hat{y}_i - y_i )^2 \label{equation:mse} \\
    \text{RMSE} =& \sqrt{\frac{1}{N} \sum_{i=1}^{N}(\hat{y}_i - y_i)^2} \label{equation:rmse}
\end{align}

\subsection{Percentile Bootstrap Formulation}
\label{appendix:equations-bootstrap}
Let the original evaluation dataset be $D = \{(y_{\text{true}, i}, y_{\text{pred}, i}) \mid i=1, \ldots, n\}$, where $n$ is the total number of evaluation samples.

\noindent We generate $B$ bootstrap samples. 
Each sample $D_b$ for $b = 1, \ldots, B$ is created by drawing a set of $n$ indices $I_b$ randomly with replacement from the original indices $\{1, \ldots, n\}$. 
This results in the bootstrap sample:
$$D_b = \{(y_{\text{true}, i}, y_{\text{pred}, i}) \mid i \in I_b\}$$
\noindent For each bootstrap sample $D_b$, our evaluation metric $\theta$ is calculated, yielding a distribution of $B$ bootstrap estimates:
$$\hat{\theta}_b = \text{Metric}(D_b) \quad \text{for } b = 1, \ldots, B$$
\noindent The set of all estimates is $\{\hat{\theta}_1, \hat{\theta}_2, \ldots, \hat{\theta}_B\}$. 
To construct a $(1 - \alpha)$ confidence interval, we first define the significance level. For our 95\% CI, $\alpha = 0.05$.
The lower and upper percentile bounds, $P_1$ and $P_2$, are calculated as:
\begin{align*}
    P_1 =& \left(\frac{\alpha}{2}\right) \times 100 = \left(\frac{0.05}{2}\right) \times 100 = 2.5 \\
    P_2 =& \left(1 - \frac{\alpha}{2}\right) \times 100 = \left(1 - \frac{0.05}{2}\right) \times 100 = 97.5
\end{align*}
\noindent The final $(1 - \alpha)$ percentile confidence interval for the metric $\theta$ is:
$$\text{CI}_{1-\alpha}(\theta) = \left[ \hat{\theta}^{(P_1)}, \hat{\theta}^{(P_2)} \right]$$
\noindent where $\hat{\theta}^{(p)}$ denotes the $p$-th percentile of the sorted list of bootstrap estimates $\{\hat{\theta}_b\}_{b=1}^B$.
\end{document}